\documentclass[11pt,cleanfoot]{ms}

\usepackage{epsfig} %% for loading postscript figures

\usepackage{epsfig}
\usepackage{graphicx}
\usepackage{fancyhdr}
\usepackage{setspace}
\usepackage{helvet}
\usepackage[hyphens]{url}
\usepackage{currvita}

\addtocontents{lof}{\cftpagenumbersoff{figure}}
\addtocontents{lot}{\cftpagenumbersoff{table}}
\makeatletter
\renewcommand\@dotsep{400}
\makeatother

\title{SPH simulations of turbulent flow in curved pipes with different
geometries. A comparison with experiments}

\author{C. E. Alvarado-Rodr\'{\i}guez 
    \affiliation{
        Direcci\'on de C\'atedras CONACYT, Av. Insurgentes Sur 1582,\\
        Cr\'edito Constructor, Benito Ju\'arez, 03940 Ciudad de M\'exico, Mexico\\
	and\\
        Departamento de Ingenier\'{\i}a Qu\'{\i}mica, DCNyE, Universidad de Guanajuato\\
        Noria Alta S/N, 36000 Guanajuato, Guanajuato, Mexico\\
        E-mail: carlos.alvarado@conacyt.mx}
}
\author{L. Di G. Sigalotti\thanks{Corresponding author.}
    \affiliation{
        Departamento de Ciencias B\'asicas, Universidad Aut\'onoma Metropolitana\\
        (UAM-A), Av. San Pablo 180, 02200 Ciudad de M\'exico, Mexico\\
        E-mail: leonardo.sigalotti@gmail.com}
}
\author{J. Klapp, C. R. Fierro-Santill\'an, F. Arag\'on
    \affiliation{
        Instituto Nacional de Investigaciones Nucleares, (ININ)\\
        Carretera M\'exico-Toluca km. 36.5, La Marquesa, 52750 Ocoyoacac\\
        Estado de M\'exico, Mexico\\
        E-mail: jaime.klapp@inin.gob.mx, celia.fierro.estrellas@gmail.com,
	micme2003@yahoo.com.mx}
}
\author{A. R. Uribe-Ram\'{\i}rez
    \affiliation{
        Departamento de Ingenier\'{\i}a Qu\'{\i}mica, DCNyE, Universidad de Guanajuato\\ 
	Noria Alta S/N, 36000 Guanajuato, Guanajuato, Mexico\\
        E-mail: agustin@ugto.mx}
}
\begin{document}

\maketitle    
\doublespacing
%%%%%%%%%%%%%%%%%%%%%%%%%%%%%%%%%%%%%%%%%%%%%%%%%%%%%%%%%%%%%%%%%%%%%%
\begin{abstract}
{\it The swirling secondary flow in curved pipes is studied in three-space dimensions
using a weakly compressible Smoothed Particle Hydrodynamics (WCSPH) formulation
coupled to new non-reflecting outflow boundary conditions. A large-eddy
simulation (LES) model for turbulence is benchmarked with existing experimental
data. After validation of the present model against experimental results for a
$90^{\circ}$ pipe bend, a detailed numerical study aimed at reproducing
experimental flow measurements for a wide range of Reynolds numbers has been
performed for different pipe geometries, including U pipe bends,
S-shaped pipes and helically coiled pipes. In all cases, the SPH calculated
behavior shows reasonably good agreement with the measurements across and
downstream the bend in terms of streamwise velocity profiles and cross-sectional
contours. Maximum mean-root-square deviations from the experimentally obtained
profiles are always less than $\sim 1.8$\%. This combined with the very good
matching between the SPH and the experimental cross-sectional contours shows the
uprising capabilities of the present scheme for handling engineering applications
with streamline curvature, such as flows in bends and manifolds.}
\end{abstract}

\begin{keywords}
Computational fluid dynamics (CFD); Pipe flow; Pipe bends; Confined turbulent flow;
Large-eddy simulation (LES); Mathematical modeling
\end{keywords}

%%%%%%%%%%%%%%%%%%%%%%%%%%%%%%%%%%%%%%%%%%%%%%%%%%%%%%%%%%%%%%%%%%%%%%
\begin{nomenclature}
\entry{${\rm Re}$}{Reynolds number}
\entry{${\rm De}$}{Dean number}
\entry{$\alpha$}{Womersley number}
\entry{$\rho$}{Mass density, kg m$^{-3}$.}
\entry{${\bf v}$}{Fluid velocity vector, m s$^{-1}$.}
\entry{${\bf g}$}{Gravitational acceleration, m s$^{-2}$.}
\entry{$p$}{Pressure, Pa.}
\entry{${\bf F}$}{Body force, m s$^{-2}$.}
\entry{$c$}{Sound speed, m s$^{-1}$.}
\entry{${\bf x}$}{Position vector, m.}
\entry{$\rho _{0}$}{Reference density, kg m$^{-3}$.}
\entry{$p_{0}$}{Reference pressure, Pa.}
\entry{$c_{0}$}{Reference sound speed, m s$^{-1}$.}
\entry{$\tilde {\bf v}$}{Mean velocity, m s$^{-1}$.}
\entry{${\bf v}^{\prime}$}{Fluctuating velocity, s$^{-1}$.}
\entry{${\bar\rho}$}{Reynolds-average density, kg m$^{-3}$.}
\entry{${\bar p}$}{Reynolds-average pressure, Pa.}
\entry{$T_{ij}$}{SPS stress tensor, kg m$^{-3}$ s$^{-1}$.}
\entry{$\tilde S_{ij}$}{Strain rate tensor, s$^{-1}$.}
\entry{$\delta _{ij}$}{Kronecker delta}
\entry{$t$}{Time, s.}
\entry{$M$}{Mach number}
\entry{$T$}{Averaging time, s.}
\entry{$x,y,z$}{Cartesian coordinates, m.}
\entry{$\nu$}{Kinematic viscosity, m$^{2}$ s$^{-1}$.}
\entry{$\nu _{t}$}{Smagorinsky eddy viscosity, m$^{2}$ s$^{-1}$.}
\entry{$\gamma$}{Adiabatic exponent.}
\entry{$r$}{Pipe radius, m.}
\entry{$d$}{Pipe diameter, m.}
\entry{$R_{\rm c}$}{Bend mean radius of curvature, m.}
\entry{$L$}{Pipe length, m.}
\entry{$P$}{Coil pitch, m.}
\entry{$\dot m$}{Mass flow rate, kg s$^{-1}$.}
\entry{$v_{\rm B}$}{Bulk flow velocity, m s$^{-1}$.}
\entry{$Q$}{Volumetric flow rate, m$^{3}$ h$^{-1}$.}
\entry{$\omega$}{Angular frequency, s$^{-1}$.}
\entry{$\Delta p$}{Pressure drop, Pa.}
\entry{$f_{\rm c}$}{Fanning friction factor}
\entry{$\theta$}{Angle}
\entry{$W$}{Kernel function, m$^{-3}$.}
\entry{$h$}{Smoothing length, mm.}
\entry{$m$}{Particle mass, kg.}
\entry{$\Delta t$}{Time step, s.}
\entry{$N$}{Total number of particles.}
\entry{$n$}{Number of neighboring particles.}
\entry{$a,b$}{Particle indices.}
\entry{${\bf v}_{wall}$}{Fluid velocity at the pipe wall, m s$^{-1}$.}
\entry{$0$}{Reference index.}
\end{nomenclature}

%%%%%%%%%%%%%%%%%%%%%%%%%%%%%%%%%%%%%%%%%%%%%%%%%%%%%%%%%%%%%%%%%%%%%%
\section{Introduction}

Flow in curved pipes and pipe bends are recurrent in many thermal engineering devices,
such as oil and gas pipeline systems, water supply systems, automotive engines, fluid
turbomachinery, power generation plants and heat exchangers, just to mention a few.
Such flows are complex in nature and often turbulent (i.e., inherently unsteady with
strong secondary flows superimposed to the streamwise flow). Therefore, in response to
the always increasing practical interest to understand the physical mechanisms underlying
turbulent flows through pipe bends and elbows, numerous experimental and numerical
investigations have been devoted to their study in curved pipes and ducts of various
cross-sectional shapes. In contrast to straight pipe flow, pressure drop and heat/mass
transfer are both enhanced in pipe bends, whereas turbulent flow motion can cause
vibrations and mechanical fatigue of the pipeline system, eventually leading to failure.

Representative experimental measurements of turbulent flow in curved pipes have been
conducted by Humphrey et al. \cite{Humphrey1981}, Taylor et al. \cite{Taylor1982},
Enayet et al. \cite{Enayet1982}, Chang et al. \cite{Chang1983}, Sreenivasan and Strykowski
\cite{Sreenivasan1983}, Azzola et al. \cite{Azzola1986}, Fiedler \cite{Fiedler1997}, Sudo
et al. \cite{Sudo1998,Sudo2000,Sudo2001}, El-Gammal et al. \cite{Gammal2010}, Hellstr\"om
et al. \cite{Hellstrom2011}, De Amicis et al. \cite{DeAmicis2014}, Mazhar et al.
\cite{Mazhar2016} and Oki et al. \cite{Oki2017} to mention a few. In particular, the
turbulent flow downstream a $90^{\circ}$ pipe bend have been investigated by Humphrey et
al. \cite{Humphrey1981}, Taylor et al. \cite{Taylor1982}, Enayet et al. \cite{Enayet1982},
Sudo et al. \cite{Sudo1998,Sudo2001}, and more recently by Kalpakli and \"Orl\"u
\cite{Kalpakli2013}, Mazhar et al. \cite{Mazhar2016} and Kalpakli et al. \cite{Kalpakli2016}.
In general, most of these studies report profiles of the longitudinal and circumferential
velocity components as well as cross-stream velocity contours both upstream the bend inlet,
inside the bend and downstream the bend outlet for a range of Reynolds
(Re) numbers. On the other hand, measurements of turbulent flow through circular- and
rectangular-sectioned U-bends have also been reported by Chang et al. \cite{Chang1983},
Azzola et al. \cite{Azzola1986} and Sudo et al. \cite{Sudo2000}. Moreover, experimental
works in helically coiled pipes were focused mainly on measuring pressure or temperature
data and deducing characteristics of the flow. In particular, Sreenivasan and Strykowski
\cite{Sreenivasan1983} studied the stabilization effects of the flow in the coiled section,
and more recently Cioncolini and Santini \cite{Cioncolini2006} and Hayamizu et al.
\cite{Hayamizu2008} studied the curvature and torsion effects on the flow regarding the
laminar to turbulent transition. On the other hand, experiments by Gupta et al.
\cite{Gupta2011}, Pimenta and Campos \cite{Pimenta2012} and De Amicis et al.
\cite{DeAmicis2014} were addressed to measure the pressure drop for helical pipes under
laminar flow conditions. Pulsatile flow in S-shaped pipes was investigated by Oki et al.
\cite{Oki2017} using conditions for the flow field resembling an automotive engine
environment.

As a fluid flows through a pipe bend or elbow, a secondary flow arises due to the
cross-sectional pressure gradient that takes place in the radial direction of the bend
curvature as a result of the centrifugal force acting on the fluid \cite{Sudo1998}. The
strength of the secondary flow depends on the radius of the bend curvature, $R_{\rm c}$,
and the Reynolds number, defined as ${\rm Re}=u_{\rm B}d/\nu$, where $u_{\rm B}$ is the
bulk velocity, $d$ is the pipe diameter and $\nu$ is the kinematic viscosity. If the
ratio $R_{\rm c}/d<1.5$, flow separation occurs immediately downstream the bend inlet,
giving rise to increased pressure losses \cite{Hellstrom2011}. Conversely, if
$R_{\rm c}/d>1.5$, a secondary flow consisting of two counter-rotating vortices forms,
the so-called Dean vortices \cite{Dean1927}. Under these conditions, the streamwise
velocity contours become C-shaped and the velocity profiles distort as they are shifted
away from the center of curvature of the bend \cite{Hellstrom2011}.

Due to its complex and resource-demanding character, numerical simulations of
flow in curved pipes have only started to appear in the last two decades in parallel
with the advent of increasing computing facilities. In-depth Computational Fluid Dynamics
(CFD) studies of water flow in a 90$^{\circ}$ elbow of circular cross-section with
$R_{\rm c}/d=1.4$ and ${\rm Re}=5.4\times 10^{5}$ were first reported by Homicz
\cite{Homicz2004} with the aid of the software FLUENT, where finite volume methods (FVM)
were employed to solve the Reynolds-Averaged Navier-Stokes (RANS) equations coupled to a
$k$-$\epsilon$ closure for the turbulence model. Further FVM calculations of water
flow in pipe bends were presented by Tanaka et al. \cite{Tanaka2009} for various values
of ${\rm Re}$ and $R_{\rm c}/d$, using the Large Eddy Simulation (LES) model for
turbulence. A database for LES and Direct Numerical Simulations (DNS) on pipe bend
flows can be found in Refs. \cite{Huttl2001,Noorani2013}. More recently, Kim et al.
\cite{Kim2014} performed numerical simulations of Sudo et al.'s \cite{Sudo1998}
experiments using the FVM-based OpenFOAM software along with a comparative analysis to
choose the turbulence model that better reproduced the experimental data. They found
that the Renormalization Group (RNG) $k$-$\epsilon$ model gave better results for
primary streamwise and secondary swirling velocity profiles than the other
turbulence models for $R_{\rm c}/d>1.5$ and ${\rm Re}$ ranging from $5\times 10^{4}$
to $2\times 10^{5}$. The OpenFOAM solver was also employed by R\"ohrig et al.
\cite{Rohrig2015} for simulations of turbulent flow in a $90^{\circ}$ pipe bend. They
performed a comparative computational assessment between the wall-resolved LES and
various RANS models for ${\rm Re}$ varying from $1.4\times 10^{3}$ to $3.4\times 10^{3}$.
By comparing their results with the experimental data of Kalpakli and \"Orl\"u
\cite{Kalpakli2013}, they assessed the superiority of LES over the applied RANS methods,
however at the cost of much greater computational effort. {\bf In particular, they
found that LES reproduced the experimental profiles for the streamwise mean velocity
and turbulent intensity just behind the elbow much better than RANS, while inside
the bend both approaches yielded comparatively similar mean velocity profiles and
pressure coefficients.} Further FVM-based calculations of turbulent flow in $90^{\circ}$ 
elbows were presented by Dutta et al. \cite{Dutta2016} and Wang et al. \cite{Wang2016}.

On the other hand, the effect of the helical geometry on the flow field was also
investigated numerically by a number of authors. In particular, Di Piazza and Ciofalo
\cite{DiPiazza2010} assessed the applicability of the classical $k$-$\epsilon$
and $k$-$\omega$ closure models to the prediction of pressure drop and heat
transfer in coiled tubes, using the numerical code ANSYS CFX 11 for a broad range of
${\rm Re}$, Prandtl numbers (${\rm Pr}$) and coil curvatures, while Jayakumar et al.
\cite{Jayakumar2010} studied the flow characteristics in helical pipes with various
derived correlations for the prediction of local values of the Nusselt number (${\rm Nu}$)
as a function of the average ${\rm Nu}$ and the angular position along the pipe
circumference. Later on, De Amicis et al. \cite{DeAmicis2014} investigated the laminar
flow in helically coiled pipes with the aid of three commercial CFD softwares, namely
the FLUENT, the OpenFOAM and the Finite Element based COMSOL Multiphysics packages. All
these simulations predicted a centrifugal force acting on the fluid as a consequence of
the duct curvature and a cross-sectional pressure gradient, which led to the
formation of two counter-rotating vortices that are symmetrical in the case of a toroidal
pipe. More recently, Tang et al. \cite{Tang2016} carried out CFD simulations to explore
the variations in the velocity distribution, pressure field and secondary flow when
varying the curvature radius, the coil pitch and the Dean number, defined as
${\rm De}={\rm Re}(r/R_{\rm c})^{1/2}$, where $r$ is the pipe radius. Moreover, numerical
simulations of oscillatory flow in curved pipes are of interest in the biomedical
sciences. For example, Glenn et al. \cite{Glenn2012} analyzed pulsatile flow in a
curved artery model, while van Wyk et al. \cite{vanWyk2015} performed simulations of
the pulsatile blood flow in a $180^{\circ}$ curved artery. An account of the
state-of-the-art research concerning turbulent flow in curved pipes, covering
experimental and numerical work, can be found in the review paper by Kalpakli Vester et
al. \cite{Kalpakli2016a}.

{\bf Numerical simulations of Poiseuille flow at low and moderate Reynolds numbers in 
straight channels and pipes using Smoothed Particle Hydrodynamics (SPH) have been 
performed by a number of authors 
\cite{Morris1997,Sigalotti2003,Basa2009,Adami2012,Ferrand2013,Meister2014}. In
particular, simulations of open-channel flows have been primarily addressed to test the
implementation of inlet, outlet and wall boundary conditions in SPH 
\cite{Federico2012,Liang2014,Shi2019}. In comparison, SPH simulations of curved pipe 
flows are correspondingly scarcer.} Among the few examples found in the literature are 
the works of Hou et al. \cite{Hou2014}, who employed SPH methods to study flow separation 
in two-dimensional, right-angled bends for different ratios of the channel widths and
different turning angles, and Alvarado-Rodr\'{\i}guez et al. \cite{Alvarado2017}, who
performed three-dimensional SPH calculations of Sudo et al.'s \cite{Sudo2001}
experiment for flow in a square-sectioned $90^{\circ}$ pipe bend and provided a
direct comparison with Rup et al.'s \cite{Rup2011} FLUENT-based simulations of the
same experiment. {\bf SPH models of supercritical flow in circular conduit bends 
have also started to appear \cite{Rosic2017}.} In this work we focus on three-dimensional 
(3D), LES models of
single-phase flow in curved pipes with the use of a weakly compressible SPH (WCSPH)
scheme \cite{Becker2007} coupled with a new formulation for modeling non-reflecting
outlet boundary conditions. The goal is to test the new outflow boundary conditions
and show the uprising capabilities of the present scheme for handling engineering
applications with streamline curvature, such as flows in bends and manifolds, and
reproducing experimental results with very good accuracy. Flow in curved pipes is
studied for different pipe geometries, including $90^{\circ}$ and U pipe bends,
S-shaped pipes and helically coiled pipes. The numerical simulations are addressed to
describe the effects of geometry on the flow through the onset of secondary flow, the
deformation of the axial velocity profile and reproduce the results of existing
calibrated experiments.

The governing equations, the SPH methodology and the implementation of the
boundary conditions are described in Section 2. The validation of the numerical
scheme is given in Section 3. Section 4 describes the numerical results
and provides direct comparison with experimental and numerical data in the literature.
In all cases, the results obtained from the SPH simulations are found to be
in good agreement with the experimental measurements. Finally, Section 5 contains
the main conclusions.

\section{Numerical methods and models}

\subsection{Governing equations}

The basic equations describing the flow of a fluid in a curved pipe are given by the
Navier-Stokes equations
\begin{eqnarray}
\frac{d\rho}{dt}&=&-\rho\nabla\cdot {\bf v},\\
\frac{d{\bf v}}{dt}&=&-\frac{1}{\rho}\nabla p+\nu\nabla ^{2}{\bf v}+{\bf g},
\end{eqnarray}
where $\rho$ is the mass density, ${\bf v}$ is the fluid velocity vector, $p$ is
the pressure, $\nu$ is the coefficient of kinematic viscosity, ${\bf g}$ is the
acceleration due to gravity and $d/dt=\partial /\partial t+{\bf v}\cdot\nabla$ is the
material time derivative. 
Equations (1) and (2) express the mass and momentum conservation laws, respectively. 
The dynamical pressure is related to the density by means of the Murnaghan-Tait 
equation \cite{Becker2007}
\begin{equation}
p=p_{0}\left[\left(\frac{\rho}{\rho _{0}}\right)^{\gamma}-1\right],
\end{equation}
where $\gamma =7$,
$p_{0}=c_{0}^{2}\rho _{0}/\gamma$, $\rho _{0}$ is a reference density and $c_{0}$ is the
sound speed at the reference density. The term $p_{0}$ governs the relative density
fluctuations $|\rho -\rho _{0}|/\rho _{0}\sim M^{2}$, where $M$ is the Mach number.
Typically $M^{2}$ is set to 0.01 in order to enforce density fluctuations within 1\%.
This is achieved by artificially setting the reference sound speed to be at least 10 times
higher than the maximum expected fluid velocity \cite{Becker2007}. {\bf This ensures
that the compressibility effects are purely acoustic so that they can be considered
to be superimposed to the main flow with almost no interaction. Therefore, a Mach number
equal to 0.1 satisfactorily enforces the fluid incompressibility condition. A very robust 
WCSPH scheme, known in the literature as $\delta$-SPH, was introduced by Antuono et al.
\cite{Antuono2010}, which stabilizes the conventional WCSPH scheme against noisy and 
oscillatory pressure fields. However, preliminary test calculations with this variant and 
the standard WCSPH formulation implemented here showed no differences in the final results 
for turbulent flow in a $90^{\circ}$ pipe elbow.}

Equations (1) and (2) are replaced by spatially-filtered equations pertinent to the
LES method. This will allow resolving the larger turbulent scales while still modeling
in an approximate manner the smaller ones. In particular, in the SPH framework
coherent turbulent structures in the fluid can be followed by implementing a sub-particle
scaling (SPS) technique \cite{Yoshizawa1986}. This is achieved by dividing the flow
velocity into its mean (or ensemble average) part and the fluctuations around it as
${\bf v}=\tilde {\bf v}+{\bf v}^{\prime}$, where the mean $\tilde {\bf v}$ is given by
the density-weighted Favre-filtering
\begin{equation}
\tilde {\bf v}=\frac{1}{\bar\rho}\frac{1}{T}\int _{t}^{t+T}\rho ({\bf x},t)
{\bf v}({\bf x},t)dt,
\end{equation}
where $T$ is an averaging time and $\bar\rho$ is the mean density given by the
conventional Reynolds-averaged density. Applying the Favre filtering to Eqs. (1) and
(2) yields the spatially filtered Navier-Stokes equations
\begin{eqnarray}
\frac{d\bar\rho}{dt}&=&-\bar\rho\nabla\cdot\tilde {\bf v},\\
\frac{d\tilde {\bf v}}{dt}&=&-\frac{1}{\bar\rho}\nabla\bar p+\frac{\nu}{\bar\rho}
\left[\nabla\cdot\left(\bar\rho\nabla\right)\right]\tilde {\bf v}+
\frac{\nu}{\bar\rho}\nabla\cdot {\bf T}+{\bf g},
\end{eqnarray}
where ${\bf T}$ is the SPS stress tensor, which in component form can be wtitten as
\begin{equation}
T_{ij}=\bar\rho\nu _{t}\left(2\tilde S_{ij}-\frac{2}{3}\tilde S_{kk}\delta _{ij}
\right)-\frac{2}{3}\bar\rho C_{I}\Delta ^{2}\delta _{ij}|\tilde S|^{2}.
\end{equation}
Here $\tilde S_{ij}$ is the Favre-filtered strain rate tensor given by
\begin{equation}
\tilde S_{ij}=\frac{1}{2}\left(\frac{\partial\tilde v_{i}}{\partial x_{j}}+
\frac{\partial\tilde v_{j}}{\partial x_{i}}\right),
\end{equation}
$C_{I}=0.00066$, $\nu _{t}=(C_{s}\Delta)^{2}|\tilde S|$, with $C_{s}=0.12$, is the
Smagorinsky eddy viscosity, $|\tilde S|=(2\tilde S_{ij}\tilde S_{ij})^{1/2}$ is the
local strain rate, $\delta _{ij}$ is the Kronecker delta and $\Delta$ is a measure of
the finite particle size. For practical purposes, $\Delta$ is set equal to the local
particle smoothing length $h$ (see below).

\subsection{SPH solver}

Equations (5) and (6) are solved numerically using a variant of DualSPHysics
\cite{Gomez2012}, an open-source CFD toolbox, which is based on the SPH methodology
\cite{Monaghan1992,Liu2010}. The SPH method is a fully Lagrangian, mesh-free scheme
for complex fluid-flow simulations, where the fluid is represented by a finite number
of discrete particles that characterize the flow attributes. Due to its Lagrangian
character, SPH tracks material history information of the system on the moving particles
which carry all field information such as position, density and velocity.

SPH consists of two steps. One is the kernel approximation in which the estimate of
a function, $f({\bf x})$, is given by the volume integral
\begin{equation}
\langle f({\bf x})\rangle =\int _{\Omega}f({\bf x}^{\prime})W(|{\bf x}-{\bf x}^{\prime}|,h)
d{\bf x}^{\prime},
\end{equation}
$\forall {\bf x}\in\Omega$, where $\Omega\subset {\bf R}^{3}$ is the spatial domain
of integration, $W$ is the kernel (or interpolation) function, which is assumed to be
spherically symmetric in its support domain $\Gamma$ ($\Gamma\subset\Omega$ with
boundary $\partial\Gamma$) and $h$ is the so-called smoothing length, which is a
dilation parameter that determines the (compact) support size of the kernel. If the
spatial domain $\Omega$ is divided into $N$ sub-domains, labeled $\Omega _{a}$, each of
which encloses a particle $a$ at position ${\bf x}_{a}\in\Omega _{a}$, the discrete
equivalent of Eq. (9) becomes
\begin{equation}
f_{a}=\sum _{b=1}^{n}f_{b}W_{ab}\Delta V_{b},
\end{equation}
where $f_{a}=\langle f({\bf x}_{a})\rangle$, $W_{ab}=W(|{\bf x}_{a}-{\bf x}_{b}|,h)$,
$\Delta V_{b}$ is the volume of particle sub-domain $\Omega _{b}$ and the summation
is taken over the $n$ neighbors of particle $a$ within the support $\Gamma$
of the kernel. {\bf A schematic drawing of the kernel in two dimensions is shown in Fig. 
1. A field variable at particle $a$ is approximated using the corresponding field 
variable of all neighboring particles $b$ within the support domain of radius $kh$, 
where $k$ is a scale parameter usually $\leq 2$ for most kernel functions. The support 
of the kernel shown in Fig. 1 as a circle is actually a sphere of radius $kh$ in three 
dimensions. Because of its compact nature, at radial distances from particle $a$ greater 
than $kh$ the kernel vanishes so that the number of neighbors $n$ is always a small 
subset of the total number of particles. The neighbors of particle $a$ can be unevenly
distributed within the support of the kernel and their instantaneous positions 
${\bf x}_{b}=(x_{b},y_{b},z_{b})$ are here defined with respect to an inertial frame 
at rest with the pipe walls.} Under the assumption of a mass density distribution 
$\rho ({\bf x})$, the particle mass, $m_{b}$, is defined as $\rho _{b}\Delta V_{b}$ so 
that Eq. (10) is finally rewritten as
\begin{equation}
f_{a}=\sum _{b=1}^{n}\frac{m_{b}}{\rho _{b}}f_{b}W_{ab}.
\end{equation}
This step is called particle (or SPH) approximation. Density fluctuations are calculated
using the continuity equation (5), which in SPH form can be written as
\begin{equation}
\frac{d\rho _{a}}{dt}=\sum _{b=1}^{n}m_{b}\left({\bf v}_{a}-{\bf v}_{b}\right)
\cdot\nabla _{a}W_{ab},
\end{equation}
where the density may be either the local density $\rho$ or the particle-scale
density $\bar\rho$ depending on whether we are dealing with laminar or turbulent
flows. The same applies to the velocity field where ${\bf v}$ may represent a local
velocity (for laminar flows) or the Favre-filtered velocity $\tilde {\bf v}$ (for
turbulent flows).

The source terms in Eqs. (2) and (6) are written in SPH form using the symmetric
representation proposed by Colagrossi and Landrini \cite{Colagrossi2003} for the
pressure gradient and the representations given by Lo and Shao \cite{Lo2002} for
the laminar viscous term and the SPS stresses. Therefore, in SPH form Eq. (6) becomes
\begin{eqnarray}
\frac{d{\bf v}_{a}}{dt}=&-&\frac{1}{\rho _{a}}\sum _{b=1}^{n}\frac{m_{b}}{\rho _{b}}
\left(p_{a}+p_{b}\right)\nabla _{a}W_{ab}\nonumber\\
&+&4\nu\sum _{b=1}^{n}m_{b}
\frac{{\bf v}_{a}-{\bf v}_{b}}{\rho _{a}+\rho _{b}}\frac{{\bf x}_{ab}\cdot\nabla _{a}W_{ab}}
{|{\bf x}_{ab}|^{2}+\epsilon ^{2}}\nonumber\\
&+&\sum _{b=1}^{n}m_{b}\left(\frac{{\bf T}_{a}}{\rho _{a}^{2}}+
\frac{{\bf T}_{b}}{\rho _{b}^{2}}\right)\cdot\nabla _{a}W_{ab}+{\bf g},
\end{eqnarray}
where ${\bf x}_{ab}={\bf x}_{a}-{\bf x}_{b}$ and $\epsilon ^{2}=0.01h^{2}$. The tilde
operator over the velocity and the bars over the density and pressure have been dropped
for simplicity. In Eqs. (12) and (13) the kernel function $W_{ab}$ is evaluated using the
symmetrization
\cite{Hernquist1989}
\begin{equation}
W_{ab}=\frac{1}{2}\left[W(|{\bf x}_{a}-{\bf x}_{b}|,h_{a})+
W(|{\bf x}_{a}-{\bf x}_{b}|,h_{b})\right],
\end{equation}
whenever $h_{a}\neq h_{b}$. Note that the SPH representation of Eq. (2) for laminar
flows can be recovered from Eq. (13) by simply dropping the SPS stress term. Particle
positions are determined by solving simultaneously the equation
\begin{equation}
\frac{d{\bf x}_{a}}{dt}={\bf v}_{a}+\frac{\beta x_{0}v_{\rm max}}{M}\sum _{b=1}^{N}m_{b}
\frac{{\bf x}_{ab}}{\left({\bf x}_{ab}\cdot {\bf x}_{ab}\right)^{3/2}},
\end{equation}
where the second term on the right-hand side is added to prevent the growth of SPH
discretization errors due to anisotropies in the distribution of particle positions
\cite{Vacondio2013}. Here $\beta =0.04$, $v_{\rm max}$ is the maximum fluid velocity,
$M$ is the total mass of the system and
\begin{equation}
x_{0}=\frac{1}{N}\sum _{b=1}^{N}\left({\bf x}_{ab}\cdot {\bf x}_{ab}\right)^{1/2},
\end{equation}
is the mean distance between particle $a$ and all other particles. The
summations in Eqs. (15) and (16) are over the total number of particles filling the
computational fluid domain.

A Wendland C$^{2}$ function is adopted as the interpolation kernel
\cite{Wendland1995,Dehnen2012}
\begin{equation}
W(q,h)=\frac{21}{2\pi h^{3}}\left(1-q\right)^{4}\left(1+4q\right),
\end{equation}
for $q\leq 1$ and zero otherwise, where $q=|{\bf x}-{\bf x}^{\prime}|/h$. These
interpolating functions are known to improve the convergence properties of SPH
and suppress the pairing instability regardless of the number of neighbors within
the kernel support \cite{Dehnen2012}. Particle velocities and positions are updated
using a temporally second-order accurate Verlet scheme for time integration of Eqs.
(13) and (15), given by the following difference formulae
\begin{eqnarray}
\rho _{a}^{n+1}&=&\rho _{a}^{n-1}+2\Delta t\left(\frac{d\rho _{a}}{dt}\right)^{n},
\nonumber\\
{\bf v}_{a}^{n+1}&=&{\bf v}_{a}^{n-1}+2\Delta t\left(\frac{d{\bf v}_{a}}{dt}
\right)^{n},\\
{\bf x}_{a}^{n+1}&=&{\bf x}_{a}^{n}+\Delta t{\bf v}_{a}^{n}+0.5\Delta t^{2}
\left(\frac{d{\bf v}_{a}}{dt}\right)^{n},\nonumber
\end{eqnarray}
where $\Delta t=t^{n+1}-t^{n}$. In order to prevent the time integration to yield
results that may diverge from the actual solution and improve the numerical
coupling between Eqs. (13) and (15) during the evolution, the above scheme is
replaced every 40 timesteps by the alternative scheme
\begin{eqnarray}
\rho _{a}^{n+1}&=&\rho _{a}^{n}+\Delta t\left(\frac{d\rho _{a}}{dt}\right)^{n},
\nonumber\\
{\bf v}_{a}^{n+1}&=&{\bf v}_{a}^{n}+\Delta t\left(\frac{d{\bf v}_{a}}{dt}
\right)^{n},\\
{\bf x}_{a}^{n+1}&=&{\bf x}_{a}^{n}+\Delta t{\bf v}_{a}^{n}+0.5\Delta t^{2}
\left(\frac{d{\bf v}_{a}}{dt}\right)^{n}.\nonumber
\end{eqnarray}
The optimal timestep calculation is guided by the following prescriptions
\begin{eqnarray}
\Delta t_{f}&=&\min _{a}\left(h|dv_{a}/dt|^{-1}\right)^{1/2},\nonumber\\
\Delta t_{v,a}&=&\max _{b}|h{\bf x}_{ab}\cdot{\bf v}_{ab}/({\bf x}_{ab}
\cdot {\bf x}_{ab}+\epsilon ^{2})|,\nonumber\\
\Delta t_{cv}&=&\min _{a}\left[h\left(c_{a}+\Delta t_{v,a}\right)^{-1}
\right],\nonumber\\
\Delta t&=&0.3\min\left(\Delta t_{f},\Delta t_{cv}\right),
\end{eqnarray}
where the maximum and minima are taken over all particles in the system,
$v_{a}=({\bf v}_{ab}\cdot{\bf v}_{ab})^{1/2}$ and $c_{a}$ is the sound speed
for particle $a$.

\subsection{Boundary conditions}

No-slip boundary conditions are implemented at the pipe walls using the method
of dynamic boundary particles developed by Crespo et al. \cite{Crespo2007,Crespo2015}.
To do so, the pipe surface is covered by a layer of uniformly-spaced external
particles, which are used to cope with the problem of kernel deficiency outside
the computational domain. A layer of particles is also employed to represent
the solid wall, which are updated using the same loop as the inner fluid
particles and so they are forced to satisfy Eqs. (13) and (15). However, they
are not allowed to move accordingly so that their initial positions and
velocities (${\bf v}_{\rm wall}={\bf 0}$) remain unchanged in time. In this way,
the presence of the wall is modeled by means of a repulsive force, which is
derived from the source of the momentum Eq. (13) and includes the effects of
compressional, viscous and gravitational froces. This force is exerted by the
wall particles on the fluid particles only when the latter get closer than a
distance $d_{a}=2h$ from the wall. Across the wall, an external particle, say
$a^{\prime}$, is assigned a mass $m_{a^{\prime}}=m_{\rm wall}$ and a density
$\rho _{a^{\prime}}=\rho _{\rm wall}$, while Neumann boundary conditions for the
pressure are ensured by setting $p_{a^{\prime}}=p_{\rm wall}$.

An inflow zone upstream the pipe inlet is used to set proper inlet boundary conditions.
In this zone, a steady inlet velocity without temporal fluctuations is assumed, which
may correspond to either a flat profile or a fully developed Hagen-Poiseuille flow.
The assumption of steady flow conditions at the inlet is justified because disturbances
there are expected to be low due to the long straight pipe segment upstream the bend
for most models. For the purposes of comparing the
numerical results with calibrated experiments in the literature, the boundary
conditions employed in the simulations are set identically with the corresponding
experimental cases. On the other hand, outflow boundary conditions are designed in
such a way that the flow is allowed to cross the pipe outlet without being
significantly reflected back \cite{Alvarado2017}. To this end an outflow zone is
defined downstream the pipe exit plane. In all simulations the lengths of the
inflow and outflow zones are chosen to be the same and equal to 5 initial uniform
particle separations in the streamwise direction. The method then allows for
anisotropic wave propagation across the outlet, where the velocity vector of
particles incident on the outlet and entering the outflow zone obeys the outgoing
wave equation \cite{Alvarado2017}
\begin{equation}
\frac{\partial {\bf v}}{\partial t}+v_{x}\frac{\partial {\bf v}}{\partial x}
-\nu\left(\frac{\partial ^{2}{\bf v}}{\partial y^{2}}+
\frac{\partial ^{2}{\bf v}}{\partial z^{2}}\right)={\bf 0},
\end{equation}
where ${\bf v}=(v_{x},v_{y},v_{z})$. This equation is valid for the case where
the streamwise flow is assumed to be along the $x$-direction. The velocity of
particles in the outflow zone is evolved using Eq. (21). For an outflow particle,
say $o$, this equation can be written in SPH form as
\begin{eqnarray}
\frac{\partial {\bf v}_{o}}{\partial t}&=&-v_{x,o}\sum _{b=1}^{n}\frac{m_{b}}
{{\bar\rho}_{ob}}\left({\bf v}_{b}-{\bf v}_{o}\right)\frac{\partial W_{ob}}
{\partial x_{o}}\nonumber\\
&+&2\nu\sum _{b=1}^{n}\frac{m_{b}}{\rho _{b}}
\frac{\left({\bf v}_{b}-{\bf v}_{o}\right)}{|{\bf x}_{ob}|^{2}+\epsilon ^{2}}
\left(y_{ob}\frac{\partial W_{ob}}{\partial y_{o}}+z_{ob}
\frac{\partial W_{ob}}{\partial z_{o}}\right),\nonumber
\end{eqnarray}
where ${\bf x}_{ob}={\bf x}_{o}-{\bf x}_{b}$, $y_{ob}=y_{o}-y_{b}$,
$z_{ob}=z_{o}-z_{b}$ and ${\bar\rho}_{ob}=(\rho _{o}+\rho _{b})/2$. Since in most
problems of interest the inlet and outlet mass rates and cross-sectional areas may
differ, a particle leaving the outflow zone is temporarily stored in a reservoir
buffer and its velocity is automatically zeroed. Only when an inflow particle
crosses the pipe inlet, a particle is removed from the buffer and placed in the
upstream side of the inflow zone with the desired prescribed density and
velocity values. The initial number of reservoir particles is arbitrarily set
and can be as large as needed. According to the usual SPH procedures, outflow
particles near the outlet may have some fluid particles as neighbors, thereby
allowing fluid information to be propagated into the outflow zone. The numerical
stability of Eq. (22) is improved when the streamwise velocity component,
$v_{x,o}$, is smoothed according to
\begin{equation}
v_{x,o}=\sum _{b=1}^{n}\frac{m_{b}}{\rho _{b}}v_{x,b}W_{ob},
\end{equation}
where particle $o$ can have neighbors of both types (i.e., fluid and outflow
particles) depending on how close it is from the pipe outlet. The position of
outflow particles is determined by solving the equation
\begin{equation}
\frac{d{\bf x}_{o}}{dt}={\bf v}_{o},
\end{equation}
which, together with Eq. (22), is integrated using the Verlet scheme described by
Eqs. (18) and (19).

\section{Convergence test and numerical validation}

In order to provide internally coherent flow conditions for the subsequent
simulations of bent pipes, a convergence test was carried out by varying the total
number of particles, $N$. The same test was also employed to validate the numerical
procedure, which consists of simulating Enayet et al.'s \cite{Enayet1982} experiments of
water flow in a $90^{\circ}$ bend of circular cross-section at various bulk Reynolds
numbers. The geometry and dimensions of the test bend are shown in Fig. 2. The pipe
diameter is $d=48$ mm and the radius of curvature is $R_{\rm c}=2.8d=134.4$ mm. As
shown in Fig. 2, a horizontal straight pipe 240 mm long and of the same diameter is
fitted upstream to the bend and a second straight pipe 480 mm long is fitted downstream.
The bulk flow velocity is defined as
\begin{equation}
v_{\rm B}=\frac{4{\dot m}}{\pi\rho d^{2}},
\end{equation}
where $\rho$ is the density of water and ${\dot m}$ is the mass flow rate given by
${\dot m}=\pi\rho\nu d{\rm Re}/4$, and $\nu$ is the kinematic viscosity of water
(taken to be $\approx 10^{-6}$ m$^{2}$ s$^{-1}$ at $20^{\circ}$C) and
${\rm Re}$ is the diameter-based bulk Reynolds number. The coordinate axes are chosen
such that the streamwise local mean velocity is always along the $x$-direction
($\tilde v=v_{x}$). Experimental results for this test case were obtained by Enayet et al.
\cite{Enayet1982} for laminar flow at ${\rm Re}=500$ and 1093 and for turbulent flow
at the maximum obtainable ${\rm Re}=43000$, corresponding to values of $v_{x}=10.5$
mm s$^{-1}$, 23 mm s$^{-1}$ and 0.92 m s$^{-1}$, respectively.

The results of the convergence study are reported in Fig. 3, where the streamwise
velocity profiles in the horizontal ($y$) and vertical ($z$) directions normalized
to $v_{\rm B}$ are shown for the various ${\rm Re}$ and resolutions considered. The
profiles for ${\rm Re}=500$, 1093 and 43000 are compared with the experimental data
of Enayet et al. \cite{Enayet1982} at 212.16 mm from the pipe inlet (upstream the
bend inlet). Table 1 lists the deviations between the experimental and numerical
profiles in terms of the root-mean-square error (RMSE). The numerical profiles in
Fig. 3 show an asymptotic tendency to globally converge as the number of particles
is increased from $N=247790$ to $4038085$. This is also seen in Table 1, where the
RMSE values decrease progressively with the number of particles. Independently of
the ${\rm Re}$-value, the RMSEs for $N=4038085$ particles are always less than
$\sim 1.6$\%. On the basis of these results, a spatial resolution resulting from
$N\geq 3000000$ particles is adopted for all subsequent simulations conducted in the
present study. {\bf Working with $N=4038085$ particles a full run was completed 
after 435698 timesteps. This took 3.03 hours of real time with an Nvidia Tesla V100 
16 GB memory computer graphics card with 5120 CUDA cores and a maximum clock rate of 
1.53 GHz.}

The generation of suitable inflow conditions in the pipe bend is further investigated
to validate the numerical procedure. Figure 4 shows intensity maps of the streamwise
velocity in four cross-stream planes, i.e., at $\theta =30^{\circ}$, $60^{\circ}$ and
$75^{\circ}$ within the bend and downstream the bend at $x=d$ (i.e., at 432 mm from
the pipe outlet for ${\rm Re}=500$ (left) and ${\rm Re}=1093$ (right). The effects of
a strong secondary flow are clearly seen, which in both cases displaces the region of
maximum velocity to the outside of the bend. These maps reproduce remarkably well the
velocity contours that were measured experimentally by Enayet et al. \cite{Enayet1982}
at the same pipe stations (see their Figs. 4 and 6). The maximum velocities are
$v/v_{\rm B}\approx 2.0$ for ${\rm Re}=500$ and $\approx 1.91$ for ${\rm Re}=1093$,
which are very close to the values quoted by Enayet et al. in their Figs. 4 and 6. As
the liquid flows into the bend, the secondary flow develops gradually. For instance,
at $\theta =30^{\circ}$ a thickening of the shear layer on the inside of the bend is
already visible, while at $\theta =60^{\circ}$ and $75^{\circ}$ the secondary flow is
almost fully developed as shown by the C-shaped contours. The resulting cross-sectional
velocity maps for the turbulent flow at ${\rm Re}=43000$ are shown in Fig. 5 at five
different stations, namely $\theta =30^{\circ}$, $60^{\circ}$ and $75^{\circ}$ within
the bend and downstream the bend at 432 mm and 288 mm from the pipe outlet (see Fig. 2),
corresponding to stations $x=d$ and $x=6d$ downstream the bend, respectively. In
the $60^{\circ}$ plane, the flow has become distorted due to the presence of a strong
secondary flow, which then grows and extends over and beyond the inner half of the
section as shown by the velocity map at $75^{\circ}$. In much the same way as in
Enayet et al.'s experiment, the large secondary flow extends over the entire flow
area and persists at 288 mm from the pipe outlet, where the C-shape deformation
is still evident. As in the laminar case, the morphology of the velocity maps is also
seen to match very well the experimental velocity contours of Enayet et al.'s (see
their Fig. 9).

{\bf Modern applications of Eq. (4) to model water at all temperatures use $\gamma =7$.
However, experimental fittings for pure and sea water data at $20^{\circ}$C yielded
values of the exponent between $\sim 5$ and $\sim 8$ \cite{MacDonald1966,Li1967}.
In order to test the level of dependence of the results on $\gamma$, we use the
test model with ${\rm Re}=43000$ and $N=4038085$ for $5\leq\gamma\leq 9$. The
results for the horizontal and vertical streamwise velocity profiles are shown in
Fig. 6. As $\gamma$ is varied the numerical profiles remain essentially the same. Table 
2 lists their deviations from the experimental data. The largest RMSEs always occur
for $\gamma =5$ and as the exponent is increased to $\gamma =9$ the errors converge
asymptotically to the same value. This demonstrates that the results are independent
of $\gamma$. The turbulent ${\rm Re}=43000$ model was also used to 
test the level of mass conservation. The variation of the total fluid mass, $M$, with
time can be calculated by direct integration of Eq. (13) over the pipe volume and then
by summing over all fluid particles to yield
\begin{equation}
M^{n+1}=M^{n}+\Delta t\sum _{a=1}^{N}\frac{m_{a}}{\rho _{a}^{n+1/2}}\left(
\sum _{b=1}^{n}m_{b}{\bf v}_{ab}^{n+1/2}\cdot\nabla _{a}^{n+1/2}W_{ab}^{n+1/2}\right),
\end{equation}
where the summation term between parentheses is just the SPH representation of 
$-{\bar\rho}\nabla\cdot\tilde {\bf v}$ in Eq. (13). Figure 7 depicts the time variation 
of the total fluid mass as calculated from Eq. (25) for all resolutions of Table 1.
Mass conservation is very well reproduced by all models independently of the spatial
resolution. During the first 0.05 s the model exhibits at all resolutions peak
deviations from the initial mass that are all within $3.4\times 10^{-5}$ kg. Thereafter,
the total mass oscillates with deviations amplitudes which are less than about
$6.3\times 10^{-6}$ kg. After 0.1 s the oscillations damp out and the total mass remains
constant in time. The level of mass conservation is measured in terms of the relative
error between the initial and final value of the total mass. At the lowest resolution
($N=247790$) the error is $\approx 1.6\times 10^{-3}$\% and decays to less than
$1.2\times 10^{-5}$\% when the resolution is raised to $N=4038085$ particles. Finally, 
Fig. 8 shows the RMSEs between the streamwise velocity profiles and the experimental data 
as a function of the Mach number for the runs of Table 1, where the Mach number is 
calculated as the ratio of the maximum streamwise velocity over the sound speed of
water at $20^{\circ}$C taken to be 1481 m s$^{-1}$. Since the maximum streamwise velocity
increases with resolution (see Fig. 3), so does the Mach number. In all cases, however, 
the flow remains subsonic and the errors decrease almost linearly with the Mach number as 
the resolution increases.}

\section{Results}

In any pipe system one would expect to find bends of various angles and sharpness.
Therefore, here we compare the results of the SPH simulations with known experimental
data for different curved pipe configurations.

\subsection{Secondary flow induced by a $90^{\circ}$ elbow}

The characteristics of the secondary flow induced by a $90^{\circ}$ elbow are
further explored numerically by simulating the more recent experiments by Kim et al.
\cite{Kim2014}. In contrast to Enayet et al. \cite{Enayet1982}, the geometry is
such that upstream the bend the flow is now vertically upward as shown in Fig. 9.
In this case, the pipe diameter is $d=50.8$ mm, with a $66d\approx 3.35$ m long
section upstream and a $180d\approx 9.14$ m long section downstream interconnected
by a $90^{\circ}$ elbow with a radius of curvature $R_{\rm c}=3d=152.4$ mm. For
this test case convergent solutions are obtained using a total number of 3684081
particles filling the entire pipe section. As for
the experiments, the SPH simulations are carried out for three different flow
conditions, corresponding to streamwise bulk velocities of $v_{\rm B}=1.0$ m s$^{-1}$
(${\rm Re}=50800$; Run 1), $v_{\rm B}=2.0$ m s$^{-1}$ (${\rm Re}=101600$; Run 2) and
$v_{\rm B}=4.0$ m s$^{-1}$ (${\rm Re}=203200$; Run 3). Kim et al. \cite{Kim2014}
also performed CFD simulations of Sudo et al.'s \cite{Sudo1998} experiment using the
OpenFOAM software with various turbulence models (see their Table 3). The
Renormalization Group (RNG) $k$-$\epsilon$ model with wall function was selected for
investigating the flow on the basis that it reproduced the Sudo et al.'s experiments
better than the other models. {\bf With 3684081 particles a run took 2.17 hours of 
real time and was completed after 276340 timesteps.}

Figures 10 and 11 show details of the horizontal ($y$) and vertical ($z$) streamwise
velocity profiles at stations S$_{2}=3.5d$, S$_{3}=10d$ and S$_{4}=50d$ downstream
the bend for Run 1 ($v_{\rm B}=1.0$ m s$^{-1}$; ${\rm Re}=50800$). The SPH solutions
(solid lines) are compared with Kim et al.'s experimental data (filled dots). As in
the experiment, at $3.5d$ from the elbow exit, the flow along the $z$-direction is
shifted upward due to the induced swirling when passing through the bend, and then is
redistributed downstream (see Fig. 11). However, as shown in Fig. 10 the profiles look
much more symmetric along the horizontal cross lines at the same stations; an
experimental feature that is also well reproduced by the present SPH simulations. A
direct inspection of Figs. 12 and 13 of Kim et al. shows that the SPH solution follows
the experimental profiles more closely than their RNG $k$-$\epsilon$ models. The
RMSE deviations of the SPH profiles from the experimental data are $\sim 1.7$\%
($x=3.5d$), $\sim 1.8$\% ($x=10d$) and $\sim 1.4$\% ($x=50d$) along the horizontal
cross line (Fig. 10) and similarly along the vertical cross line. At $3.5d$ from the
elbow, the horizontal ($y$) profile is such
that the streamwise velocity is a little augmented near the pipe wall due to swirling
effects which tend to reduce the velocity in the central region. However, at $10d$ from
the bend, these velocity peaks has already been diffused by turbulent dissipation.
The dependence of the flow downstream the elbow on the Reynolds number is depicted in
Fig. 12, where the SPH simulations for Run 1, 2 and 3 are compared along the horizontal
cross lines at the same pipe stations of Figs. 10 and 11. The similarity in the structure
of the flows after the elbow is very well reproduced by the SPH calculations. As
${\rm Re}$ increases, the velocity at the center of the pipe and far from the elbow
(at $50d$) decreases very slightly. This feature is also very well reproduced by the
SPH simulations. This happens because the flattening of the velocity profile increases
with the Reynolds number in a fully developed turbulent pipe flow \cite{Zagarola1997}.

\subsection{Turbulent flow in a U-bend of circular cross-section}

SPH simulations of the turbulent flow in a U-bend pipe of circular cross-section
are now compared with the laser-Doppler measurements of the longitudinal velocity
components reported by Azzola et al. \cite{Azzola1986}. The geometry and dimensions
of the strongly curved $180^{\circ}$ pipe is shown in Fig. 13. The pipe diameter is
$d=4.45$ cm and the bend mean radius of curvature is
$R_{\rm c}=(r_{0}+r_{1})/2=3.375d\approx 15.019$ cm. The U-bend is attached to its
$\theta =0^{\circ}$ inlet and $\theta =180^{\circ}$ outlet planes by two straight
pipes of the same diameter and length equal to $54.7d\approx 184.6125$ cm. The
stations upstream (S$_{1}=2d=8.9$ cm and S$_{2}=d=4.45$ cm), within ($\theta =3^{\circ}$,
$45^{\circ}$, $90^{\circ}$, $175^{\circ}$ and $177^{\circ}$) and downstream
(S$_{3}=d=4.45$ cm, S$_{4}=2d=8.9$ cm, S$_{5}=3d=13.35$ cm, S$_{6}=4d=17.8$ cm and
S$_{7}=5d=22.25$ cm) the U bend, where the SPH profiles and cross-sectional contour
maps of the streamwise velocity component are compared with Azzola et al.'s data are
marked by segment lines normal to the pipe outer wall. As for the experiments, the
SPH simulations are performed for two different flow conditions, corresponding to
mean bulk velocities $v_{\rm B}=1.29$ m s$^{-1}$ (${\rm Re}=57400$; Run 1) and
$v_{\rm B}=2.47$ m s$^{-1}$ (${\rm Re}=110000$; Run 2). The associated Dean numbers,
${\rm De}={\rm Re}(d/R_{\rm c})^{1/2}$, are ${\rm De}=31300$ and 59900, respectively.
Good convergence is already observed for a total of $N=9200853$ particles filling
the full pipe section. {\bf At this resolution a run took 1.29 hours of real time
and needed 158960 timesteps to be completed.}

Only streamwise velocity profiles for ${\rm Re}=57400$ are depicted in Azzola et al.'s
\cite{Azzola1986} paper (their Fig. 2). They argued that except for relatively small
differences, the experimental profiles obtained for ${\rm Re}=110000$ remained
essentially the same as for ${\rm Re}=57400$. While this observation is in agreement
with the findings of Kim et al. \cite{Kim2014} for varying high ${\rm Re}$-values in
the range between 50000 and 200000, the same experimenal trends are predicted by the
SPH simulations. Figure 14 shows the streamwise velocity profiles for ${\rm Re}=57400$
(solid gray lines) and ${\rm Re}=110000$ (dashed lines) as compared with the
experimental data of Azzola et al.'s for ${\rm Re}=57400$ (empty circles) at the pipe
stations marked in Fig. 13. The agreement between the SPH and Azzola et al.'s data
is very good. The RMSE deviations between the SPH profiles and the experimental
measurements vary from $\sim 0.8$\% to $\sim 1.7$\%. In all pipe stations, the SPH
streamwise velocity profiles for ${\rm Re}=110000$ are essentially the same as those
obtained for ${\rm Re}=57400$, in good agreement with the experimental observations.
A picture of the computed flow development in the U-bend is shown in Fig. 15,
where cross-stream velocity intensity maps are depicted. The flow morphology reported by
Azzola et al. in their Fig. 5 is very well reproduced by the SPH simulations at all pipe
stations. In particular, at
$\theta =90^{\circ}$ and $135^{\circ}$ the maximum streamwise velocity is displaced
off-axis, while at $\theta =177^{\circ}$ the velocity intensity is greater toward
the outer wall of the bend, which is a clear effect of the reversal of the secondary
flow on the mid-plane. The intensity maps at the downstream station
S$_{7}$ shows that the streamwise flow experiences a too slow recovery from the adverse
pressure gradient that the fluid on the inside of the bend encounters at the bend exit.

\subsection{Flow in a helically coiled pipe}

It is well-known that helical geometry imposes a number of distinctive phenomena in the
field flow compared to the most commonly studied case of straight pipes. Similar to
the usual pipe elbows, the duct curvature produces a centripetal force on the circulating
fluid, which gives rise to a cross-sectional pressure gradient toward the external bend
of the pipe accompanied by an intense recirculation in the form of two counter-rotating
vortices \cite{Jayakumar2010,DeAmicis2014}. Here we describe three-dimensional SPH
simulations of water flow in a vertically oriented helical pipe by adopting the same
parameters employed in Tang et al.'s \cite{Tang2016} experiments and CFD calculations.
The geometry of the helically coiled pipe is shown schematically in Fig. 16, which is
described by the pipe diameter $d=2r$, the coil (or curvature) radius $R_{\rm c}$, the
coil pitch $P=2\pi b$ and the pipe length $L=2\pi R_{\rm c}$. For the present analysis
$d=61.4$ mm, $R_{\rm c}=662.5$ mm, $P=75$ mm and $L=4.163$ m, corresponding to model
Pipe 1 of Tang et al. \cite{Tang2016}. At the pipe inlet a flat axial velocity profile
is specified, corresponding to a volumetric flow rate
\begin{equation}
Q=\frac{\dot m}{\rho}=\frac{\pi}{4}v_{\rm B}d^{2}.
\end{equation}
The initial conditions for eleven selected flow models are detailed in Table 3.
Starting from the first column, Table 3 lists the flow rate, the corresponding bulk
velocity, the Reynolds number and the Dean number, respectively. These values of
$Q$ are within the range of values consistent with Tang et al.'s experimental data.
All runs for this test case were carried out with 3348466 initially uniformly
distributed particles. {\bf A typical run in Table 2 took 0.93 hours of real time
to complete 49170 timesteps.}

Figure 17 shows the dependence of the pressure drop, $\Delta p$, on the flow rate, $Q$,
as calculated with SPH (solid line) for the models listed in Table 3. The SPH results
are compared with the Tang et al.'s experimental data (empty circles) and with Ito's 
\cite{Ito1959} (dashed line) and Murakami et al.'s \cite{Murakami1971} (dottedline) 
correlations for single-phase, turbulent flow in curved pipes given by
\begin{equation}
f_{\rm c}=\frac{1}{4}\left(\frac{r}{R_{\rm c}}\right)^{1/2}\Bigg\{0.029+0.304
\left[\left(\frac{r}{R_{\rm c}}\right)^{2}{\rm Re}\right]^{1/4}\Bigg\},
\end{equation}
and
\begin{equation}
f_{\rm c}=0.079{\rm Re}^{-1/4}\left[{\rm Re}\left(\frac{r}{R_{\rm c}}\right)^{2}
\right]^{1/20},
\end{equation}
respectively. Here $f_{\rm c}$ is the Fanning friction factor, which contributes to
the pressure drop as
\begin{equation}
\Delta p=\frac{2L}{d}f_{\rm c}\rho v^{2},
\end{equation}
where $L$ is the pipe length. The SPH calculations show an acceptably good agreement
with the experimental data. In terms of the RMSE metric, the deviation between the
SPH and the experimentally obtained pressure drop is less than $1.6$\%. For comparison,
Tang et al. reported a deviation between their simulated results and the experimental
measurements of $2.9$\%. Moreover, the deviations between the two empirical calculations
and the SPH results are lower than $2.6$\%, compared to the less than $5$\% deviations
reported by Tang et al.
Streamwise velocity intensity maps are shown in Fig. 18 at various cross-sectional
planes along the pipe for the case when $Q=10$ m$^{3}$ h$^{-1}$ for ${\rm Re}=644.01$
and ${\rm De}=138.63$. In this figure the sectional planes are identified by the
angle $\theta$ from $0^{\circ}$ (pipe inlet) to $360^{\circ}$ (pipe outlet). For the
same cross-sections, the intensity maps are seen to reproduce very well the velocity
contours reported by Tang et al. \cite{Tang2016} in their Fig. 7. A maximum flow
velocity is achieved approximately at $\theta =60^{\circ}$. Unsteady flow develops
along the circumference of the coiled pipe and the area of the region of maximum
velocity reduces as the flow approaches the pipe outlet.

\subsection{Pulsatile flow in an S-shaped exhaust pipe}

SPH simulations of pulsatile flow in an S-shaped exhaust pipe of an automotive
engine are now carried out under the same conditions of the experiments conducted
by Oki et al. \cite{Oki2017}. Studying the flow conditions in such manifolds is
important for maintaining the high performance of close-coupled catalytic converters,
which are used as emission control devices in automotive engines. Such exhaust gas
flows are in general complex because of their unsteadiness and highly turbulent nature
\cite{Jeong2014}. For example, the conversion efficiency and durability of the converter
are both increased when a uniform flow is established.

Figure 19 shows a schematic of the S-shaped exhaust pipe. The pipe has a
32 mm $\times$ 32 mm square section and two $57.7^{\circ}$ bends with a radius of
curvature $R_{\rm c}=50$ mm. The instantaneous velocity which is supplied to the
pipe inlet is displayed in Fig. 20 and closely follows that given by
Oki et al. \cite{Oki2017} in their Fig. 8, corresponding to two engine cycles of
period 0.08 s each in the case of 1500 rpm. The pulsation period is 0.02 s, which is a
quarter of the period per engine cycle. For this test, the three dimensionless parameters
of importance are the Reynolds number, the Dean number and the Womersley number
\cite{Womersley1955} defined as
\begin{equation}
\alpha =\frac{d}{2}\left(\frac{\omega}{\nu}\right)^{1/2},
\end{equation}
where $\omega$ is the angular frequency of pulsation. This number is a measure of the
interaction between the inertial effects due to the pulsation and the viscous effect.
The calculated flow parameters are ${\rm Re}=48000$, ${\rm De}=27200$ and
$\alpha =70.9$. As shown in Fig. 19, streamwise velocity profiles and cross-sectional
contours are obtained numerically at S$=25.2$ mm downstream the second bend outlet,
where Oki et al.'s experimental measurements were provided. In this case, the pipe
volume was filled with 3984368 particles. {\bf With this resolution the simulation
took 0.97 hours of real time to complete 57404 timesteps.}

Figure 21 shows streamwise velocity maps along the pipe bends. In agreement with the
experimental data, the flow velocity is higher in a region around the inner wall
side of the bends as a consequence of the pressure becoming lower in these regions
\cite{Sudo1998}. A comparison with Oki et al.'s Fig. 9(b) and (d) shows that the
SPH simulations reproduce quite well the experimental data as obtained from their
two-dimensional two-component (2D2C) PIV measurements. The SPH results for the
cross-sectional streamwise flow velocity after passing through the second bend at
station S are shown in Fig. 22. A comparison with the experimental counterpart (see
Fig. 10(a) of Oki et al.) shows that the SPH simulations are reproducing very well
the details of the flow. Figure 23 compares the numerically
calculated streamwise velocity profile (solid line) with the experimental data as
obtained from their 2D2C PIV (empty squares) and stereoscopic (2D3C) PIV measurements
(empty circles). The SPH results fit very well the experimental data with maximum RMSE
deviations of $0.8$\% (for 2D2C PIV) and $1.2$\% (for 2D3C PIV).
Finally, Fig. 24 displays streamwise (intensity maps) velocity fields at the (a)
first bend inlet plane, (b) first bend outlet plane, (c) second bend inlet plane and
(d) second bend outlet plane. As in the CFD results of Oki et al. \cite{Oki2017}
(see their Fig. 15), at the entrance of the first bend the steamwise velocity is
higher toward the inner side wall of the pipe. The same is true at the exit of
the first bend. As in the first bend, at the entrance of the second bend the flow velocity 
is higher toward the inner side wall and, due to the cross-sectional pressure gradient in 
the radial direction of curvature, a well-marked C-shaped profile forms which is still
evident at the second bend exit.

\section{Concluding remarks}

Experimental data, concerning the flow characteristics within curved pipes for
a wide range of Reynolds numbers (${\rm Re}$) and different pipe geometries,
were numerically reproduced in this article with the aid of a weakly compressible
Smoothed Particle Hydrodynamics (WCSPH) method coupled to new non-reflecting outflow
boundary conditions. A large-eddy simulation (LES) model was employed for the SPH
calculation of turbulent flow in the pipe bends. The numerical methodology was
validated against Enayet et al.'s \cite{Enayet1982} experimental data for
flow in a $90^{\circ}$ pipe bend for $500\leq {\rm Re}\leq 43000$.

A quantitative assessment of the SPH methodology was carried out by comparing the
numerical results with experimental and CFD data of turbulent flow in a $90^{\circ}$
pipe bend for $50800\leq {\rm Re}\leq 203000$ \cite{Kim2014}, in a U pipe bend for
$57400\leq {\rm Re}\leq 110000$ \cite{Azzola1986}, in a helically coiled pipe for
$322\leq {\rm Re}\leq 831.8$ \cite{Tang2016} and in an S-shaped exhaust pipe of
square cross-section for pulsatile flow at ${\rm Re}=48000$ \cite{Oki2017}. The
assessment showed that the present model is able to reproduce the experimental flow
data for primary streamwise and secondary swirling velocity fields along curved pipes
of different geometry with sufficiently good accuracy for a wide range of {\rm Re}
values. Maximum root-mean-square (RMSE) deviations between the experimental and
numerical streamwise velocity profiles were always found to be within $\sim 1.8$\%.
Cross-sectional contour velocity plots upstream, along  and downstream the pipe
bends were also quantitatively reproduced by the SPH simulations of the various
cases. In particular, the contour shapes and the swirl intensities displayed
very similar characteristics to the experiments. The agreement of the flow
characteristics with the experimental data shows the uprising capabilities of the
present SPH scheme for handling engineering applications involving complicated
swirling flows in bends and manifolds.

\begin{acknowledgment}
C.E.A.-R. is a research fellow commissioned to the University of Guanajuato
(under Project No. 368) and he thanks financial support from CONACYT under this
project. The calculations of this paper were performed using computational facilities
at the Barcelona Supercomputer Center and ABACUS-Laboratorio de Matem\'atica Aplicada y
C\'omputo de Alto Rendimiento of Cinvestav. We acknowledge funding from the European
Union's Horizon 2020 Programme under the ENERXICO Project (Grant agreement No. 828947)
and under the Mexican CONACYT-SENER-Hidrocarburos (Grant agreement No. B-S-69926).
\end{acknowledgment}

\bibliographystyle{ms}

\bibliography{ms}

\clearpage

%------------------------------------------------------------------Table 1
\begin{table}[t]
\caption{Parameters and results of convergence tests}
\begin{center}
\label{tab:1}       % Give a unique label
\begin{tabular}{clll}
\\
\hline
$N$ & ${\rm Re}$ & RMSE$(v_{x_{y}}/v_{\rm B})$ & RMSE$(v_{x_{z}}/v_{\rm B})$ \\
\hline
247790 & 500 & 0.076 & 0.084 \\
~~~~~~~ & 1093 & 0.095 & 0.088 \\
~~~~~~~ & 43000 & 0.081 & 0.081 \\
523493 & 500 & 0.043 & 0.041 \\
~~~~~~~ & 1093 & 0.051 & 0.045 \\
~~~~~~~ & 43000 & 0.046 & 0.050 \\
1080924 & 500 & 0.023 & 0.021 \\
~~~~~~~ & 1093 & 0.022 & 0.018 \\
~~~~~~~ & 43000 & 0.022 & 0.035 \\
2104126 & 500 & 0.017 & 0.013 \\
~~~~~~~ & 1093 & 0.005 & 0.007 \\
~~~~~~~ & 43000 & 0.016 & 0.020 \\
4038085 & 500 & 0.015 & 0.012 \\
~~~~~~~ & 1093 & 0.004 & 0.006 \\
~~~~~~~ & 43000 & 0.015 & 0.016 \\
\hline
\end{tabular}
\end{center}
\end{table}
%-----------------------------------------------------------------------------------

\clearpage

%------------------------------------------------------------------Table 2
\begin{table}[t]
\caption{Results of the $\gamma$-independence test in Eq. (4) for 
Re$=43000$ and $N=4038085$.}
\begin{center}
\label{tab:2}       % Give a unique label
\begin{tabular}{cll}
\\
\hline
$\gamma$ & RMSE$(v_{x_{y}}/v_{\rm B})$ & RMSE$(v_{x_{z}}/v_{\rm B})$ \\
\hline
5 & 0.017 & 0.023 \\
6 & 0.015 & 0.018 \\
7 & 0.015 & 0.016 \\
8 & 0.015 & 0.017 \\
9 & 0.015 & 0.017 \\
\hline
\end{tabular}
\end{center}
\end{table}
%------------------------------------------------------------------------------

\clearpage

%------------------------------------------------------------------Table 3
\begin{table}[t]
\caption{Initial conditions for the helically coiled pipe flows}
\begin{center}
\label{tab:3}
\begin{tabular}{clll}
\\
\hline
$Q$ & $v_{\rm B}$ & Re & De \\
(m$^{3}$ h$^{-1}$) & (m s$^{-1}$) & & \\
\hline
5 & 0.469 & 322.01 & 69.32 \\
10 & 0.938 & 644.01 & 138.63 \\
15 & 1.407 & 966.02 & 207.95 \\
20 & 1.876 & 1288.02 & 277.27 \\
25 & 2.345 & 1610.03 & 346.58 \\
30 & 2.814 & 1932.03 & 415.90 \\
35 & 3.283 & 2254.04 & 485.22 \\
40 & 3.753 & 2576.04 & 554.53 \\
45 & 4.222 & 2898.05 & 623.85 \\
50 & 4.691 & 3220.05 & 693.17 \\
60 & 5.629 & 3864.06 & 831.80 \\
\hline
\end{tabular}
\end{center}
\end{table}
%-----------------------------------------------------------------------

\clearpage

\begin{figure*}
\caption{Graphical representation of the SPH approximation in two-space dimensions.
The circle of radius $kh$ defines the kernel support, which in three-space
dimensions corresponds to a sphere of radius $kh$. Only neighbor particles, $b$,
within the kernel support are used to approximate the field variables at particle,
$a$.}
\label{fig1}
\end{figure*}

\begin{figure*}
\caption{$90^{\circ}$ pipe geometry and dimensions used for the simulations of
Enayet et al.'s \cite{Enayet1982} experiments. The line segments normal to the
pipe boundary mark the stations where the streamwise velocity profiles and
cross-sectional maps are compared with the experimental data.}
\label{fig2}
\end{figure*}

\begin{figure*}
\caption{Horizontal (left column) and vertical (right column) streamwise velocity
profiles upstream the bend at 212.16 mm from the pipe inlet for ${\rm Re}=500$
(top), ${\rm Re}=1093$ (middle) and ${\rm Re}=43000$ (bottom). The SPH results
at different resolutions are compared with Enayet et al.'s \cite{Enayet1982}
experimental data (empty circles). As the resolution is increased from $N=247790$
(dashed lines) to $N=4038085$ particles (solid lines) the numerical solutions
asymptotically approach the experimental profiles.}
\label{fig3}
\end{figure*}

\begin{figure*}
\caption{Cross-sectional streamwise velocity maps at different stations
($\theta =30^{\circ}$, $60^{\circ}$ and $75^{\circ}$) within the pipe bend and
at $x=d$ downstream the bend (i.e., at 432 mm from the pipe outlet) for
${\rm Re}=500$ (right) and ${\rm Re}=1093$ (left). The inside and outside of the
pipe bend correspond to the bottom and top part of the circular maps, respectively.
The color bars indicate the magnitude of the streamwise velocity in m s$^{-1}$.}
\label{fig4}
\end{figure*}

\begin{figure*}
\caption{Cross-sectional streamwise velocity maps at different stations
($\theta =30^{\circ}$, $60^{\circ}$ and $75^{\circ}$) within the pipe bend and
at $x=d$ and $x=6d$ downstream the bend (i.e., at 432 mm and 288 mm from the
pipe outlet) for ${\rm Re}=43000$. The inside and outside of the pipe bend
correspond to the bottom and top part of the circular maps, respectively. The
color bar indicates the magnitude of the streamwise velocity in m s$^{-1}$.}
\label{fig5}
\end{figure*}

\begin{figure*}
\caption{Horizontal (top) and vertical (bottom) streamwise velocity profiles
upstream the bend at 212.16 mm from the pipe inlet for ${\rm Re}=43000$,
$N=4038085$ particles and varying $\gamma$ in Eq. (4) between 5 and 9. The
numerical profiles are compared with Enayet et al.'s \cite{Enayet1982}
experimental data (empty circles). The numerical profiles converge to the
experimental one independently of $\gamma$.}
\label{fig6}
\end{figure*}

\begin{figure*}
\caption{Time variation of the total fluid mass for ${\rm Re}=43000$ and
varying spatial resolution. The level of mass conservation improves with
resolution.}
\label{fig7}
\end{figure*}

\begin{figure*}
\caption{Deviations of the streamwise velocity profiles in terms of the
root-mean-square errors (RMSEs) as a function of the Mach number for all
runs of Table 1.}
\label{fig8}
\end{figure*}

\begin{figure*}
\caption{$90^{\circ}$ pipe geometry and dimensions used for the simulations of
Kim et al.'s \cite{Kim2014} experiments. The line segments normal to the
pipe boundary mark the stations where the streamwise velocity profiles are
compared with the experimental data.}
\label{fig9}
\end{figure*}

\begin{figure*}
\caption{Horizontal (along the $y$-direction) streamwise velocity profiles for
Run 1 with ${\rm Re}=50800$ at stations S$_{2}=3.5d$, S$_{3}=10d$ and
S$_{4}=50d$ downstream the bend exit plane as shown in Fig. 9. The SPH profiles
(solid lines) are compared with Kim et al.'s \cite{Kim2014} experimental
profiles (filled dots).}
\label{fig10}
\end{figure*}

\begin{figure*}
\caption{Vertical (along the $z$-direction) streamwise velocity profiles for
Run 1 with ${\rm Re}=50800$ at stations S$_{2}=3.5d$, S$_{3}=10d$ and
S$_{4}=50d$ downtream the bend exit plane as shown in Fig. 9. The SPH profiles
(solid lines) are compared with Kim et al.'s \cite{Kim2014} experimental
profiles (filled dots).}
\label{fig11}
\end{figure*}

\begin{figure*}
\caption{Dependence of the horizontal (along the $y$-direction) streamwise
velocity profiles on the Reynolds number. The SPH profiles for ${\rm Re}=50800$
(solid lines) ${\rm Re}=101600$ (empty circles) and ${\rm Re}=203200$ (filled
squares) are compared at pipe stations S$_{2}=3.5d$, S$_{3}=10d$ and S$_{4}=50d$
downtream the bend exit plane as shown in Fig. 9.}
\label{fig12}
\end{figure*}

\begin{figure*}
\caption{$180^{\circ}$ pipe geometry and dimensions used for the simulations of
Azzola et al.'s \cite{Azzola1986} experiments. The line segments normal to the
pipe boundary mark the stations where the streamwise velocity profiles are
compared with the experimental data.}
\label{fig13}
\end{figure*}

\begin{figure*}
\caption{Streamwise velocity profiles at different stations upstream, within and
downstream the U-bend as marked in Fig. 13. The SPH profiles (solid gray lines) are
compared with Azzola et al.'s \cite{Azzola1986} experimental profiles (empty circles)
for ${\rm Re}=57400$. For comparison the SPH profiles for ${\rm Re}=110000$ (dashed
lines) are also displayed.}
\label{fig14}
\end{figure*}

\begin{figure*}
\caption{Cross-sectional streamwise velocity maps at different stations
($\theta =3^{\circ}$, $45^{\circ}$, $90^{\circ}$, $135^{\circ}$ and $177^{\circ}$)
within the pipe bend and at S$_{3}$ ($x=d$), S$_{5}$ ($x=3d$) and S$_{7}$ ($x=5d$)
downstream the U-bend for ${\rm Re}=57400$. The inside and outside of the pipe bend
correspond to the bottom and top part of the circular maps, respectively. The
color bar indicates the magnitude of the streamwise velocity in m s$^{-1}$.}
\label{fig15}
\end{figure*}

\begin{figure*}
\caption{Schematic drawing of the helical pipe and geometrical parameters
for the simulations of Tang et al. \cite{Tang2016} experiments.}
\label{fig16}
\end{figure*}

\begin{figure*}
\caption{Dependence of the pressure drop, $\Delta p$, on the flow rate, $Q$,
for the models listed in Table 3. The SPH results (solid line) are compared
with the experimental Tang et al.'s \cite{Tang2016} data (empty circles) and
the experimental correlations derived by Ito \cite{Ito1959} (dashed line) and
Murakami et al. \cite{Murakami1971} (dotted line).}
\label{fig17}
\end{figure*}

\begin{figure*}
\caption{Streamwise velocity maps at selected cross-sectional planes along the
circumference of the helically coiled pipe for flow with $Q=10$ m$^{3}$ h$^{-1}$,
${\rm Re}=644.01$ and ${\rm De}=69.32$. The color bar indicates the magnitude of
the streamwise velocity in m s$^{-1}$.}
\label{fig18}
\end{figure*}

\begin{figure*}
\caption{Geometry and dimensions of the S-shaped exhaust pipe used in the
SPH simulations of Oki et al.'s \cite{Oki2017} experiments.}
\label{fig19}
\end{figure*}

\begin{figure*}
\caption{Instantaneous velocity at the inlet section of the S-shaped exhaust pipe
used in the SPH simulations. Figure adapted from Oki et al. \cite{Oki2017}.}
\label{fig20}
\end{figure*}

\begin{figure*}
\caption{Streamwise velocity map on a two-dimensional slice at the symmetry ($z=0$)
plane in the first (left) and second bend (right).}
\label{fig21}
\end{figure*}

\begin{figure*}
\caption{Cross-sectional streamwise velocity map downstream the second bend at
station S as shown in Fig. 19.}
\label{fig22}
\end{figure*}

\begin{figure*}
\caption{SPH streamwise velocity profiles (solid line) at station S in Fig. 19 as
compared with Oki et al.'s \cite{Oki2017} 2D2C PIV (empty squares) and 2D3C PIV
(empty circles) experimental measurements.}
\label{fig23}
\end{figure*}

\begin{figure*}
\caption{Cross-sectional streamwise velocity fields as obtained from
the SPH simulations in: (a) the first bend inlet plane, (b) the first bend outlet
plane, (c) the second bend inlet plane and (d) the second bend outlet plane. The
position of the first and second bend inlet and outlet planes are indicated in Fig.
19 by the dashed lines.}
\label{fig24}
\end{figure*}

\clearpage

\begin{figure*}
\centerline{\epsfig{figure=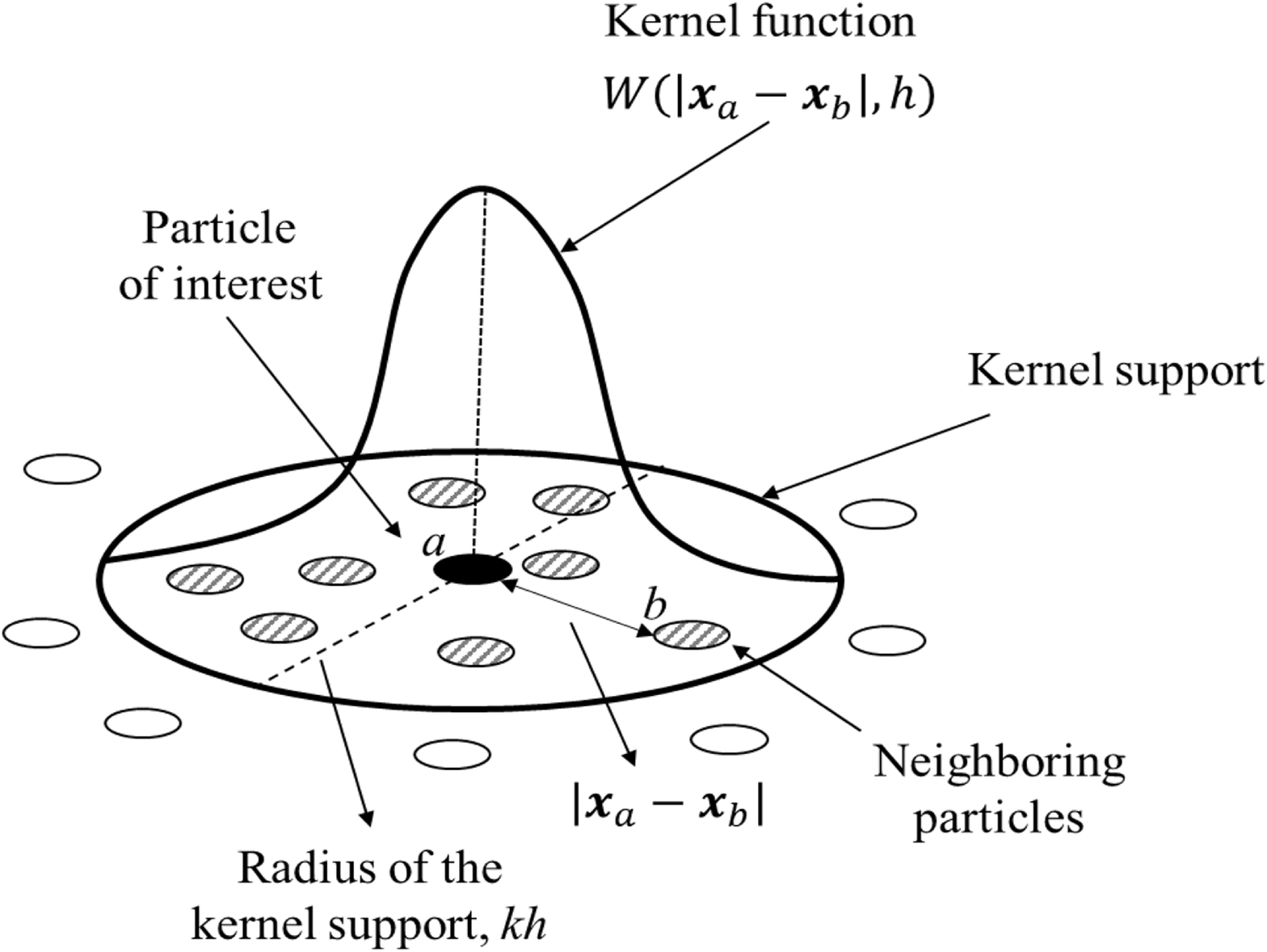,width=6.0in}}
\label{fig1}
FIGURE 1
\end{figure*}

\clearpage

\begin{figure*}
\centerline{\epsfig{figure=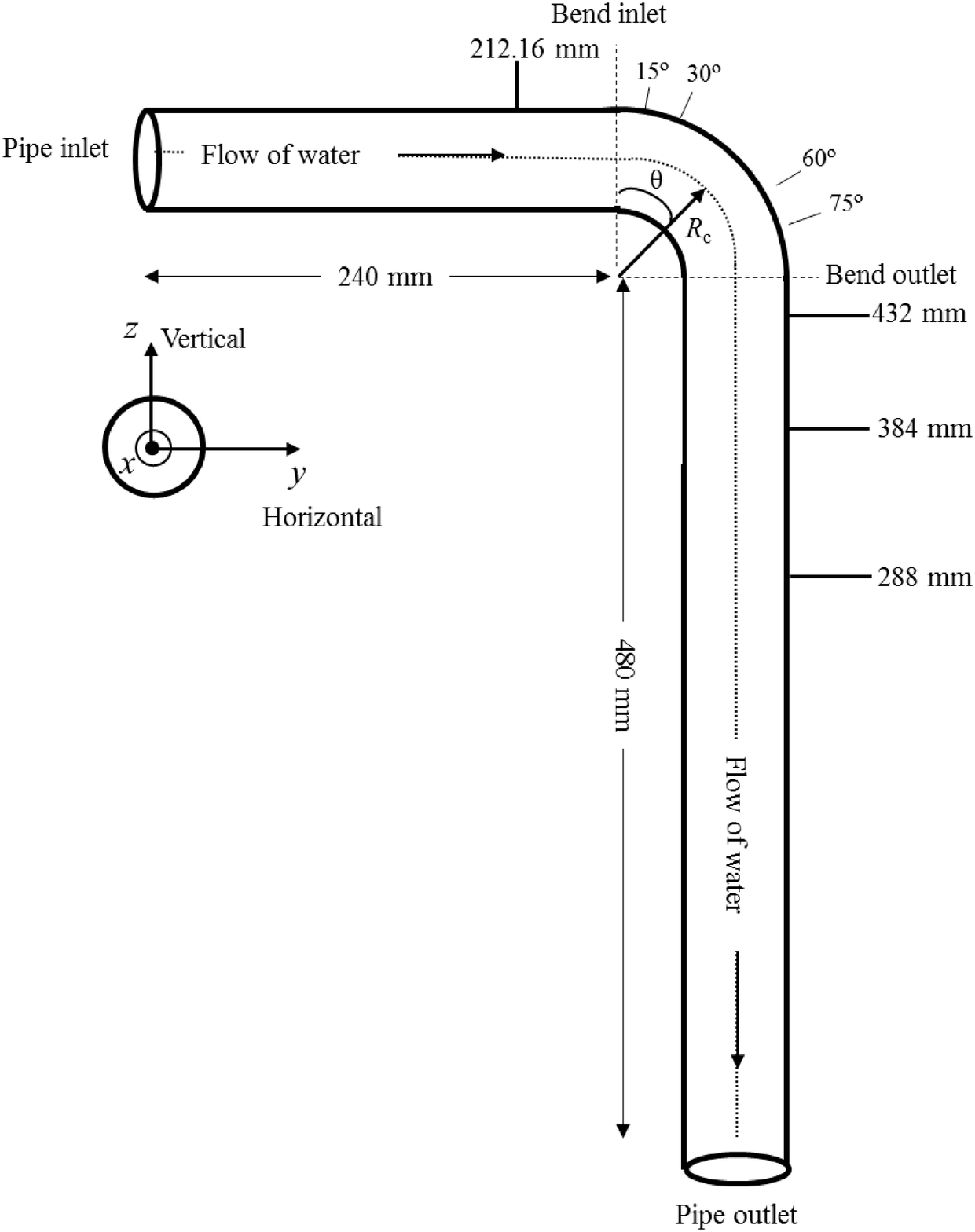,width=10.0in}}
\label{fig2}
FIGURE 2
\end{figure*}

\clearpage

\begin{figure*}
\centerline{\epsfig{figure=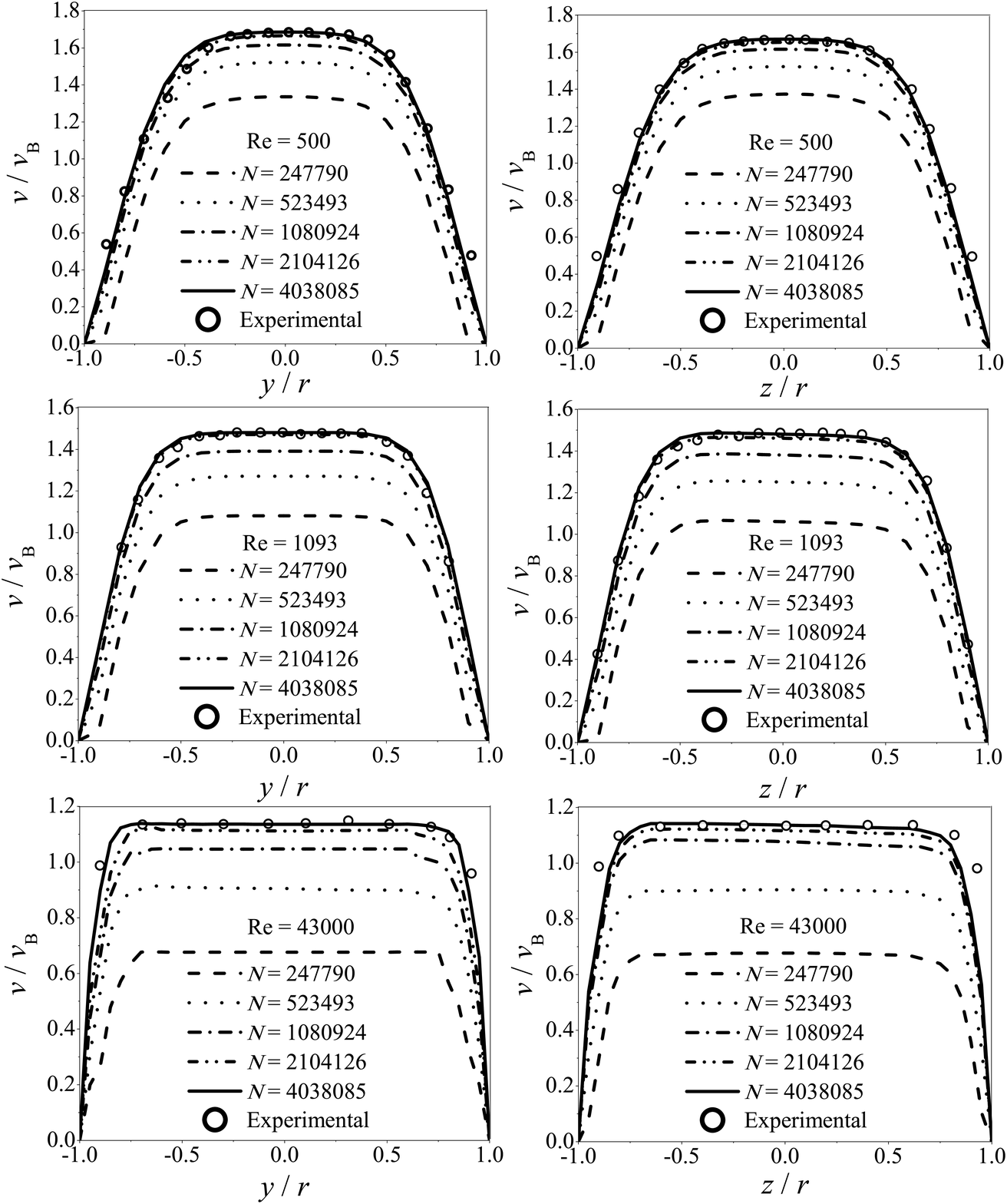,width=5.0in}}
\label{fig3}
FIGURE 3
\end{figure*}

\clearpage

\begin{figure*}
\centerline{\epsfig{figure=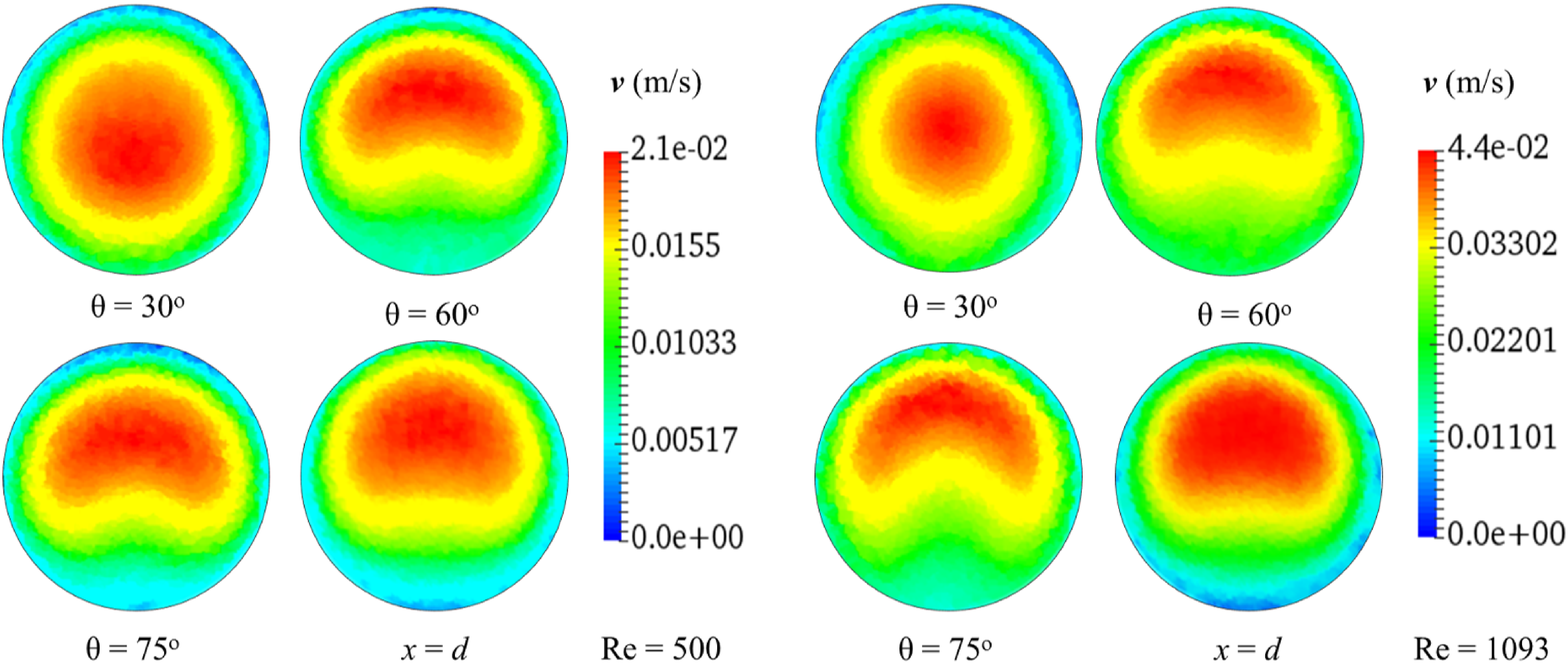,width=6.0in}}
\label{fig4}
FIGURE 4
\end{figure*}

\clearpage

\begin{figure*}
\centerline{\epsfig{figure=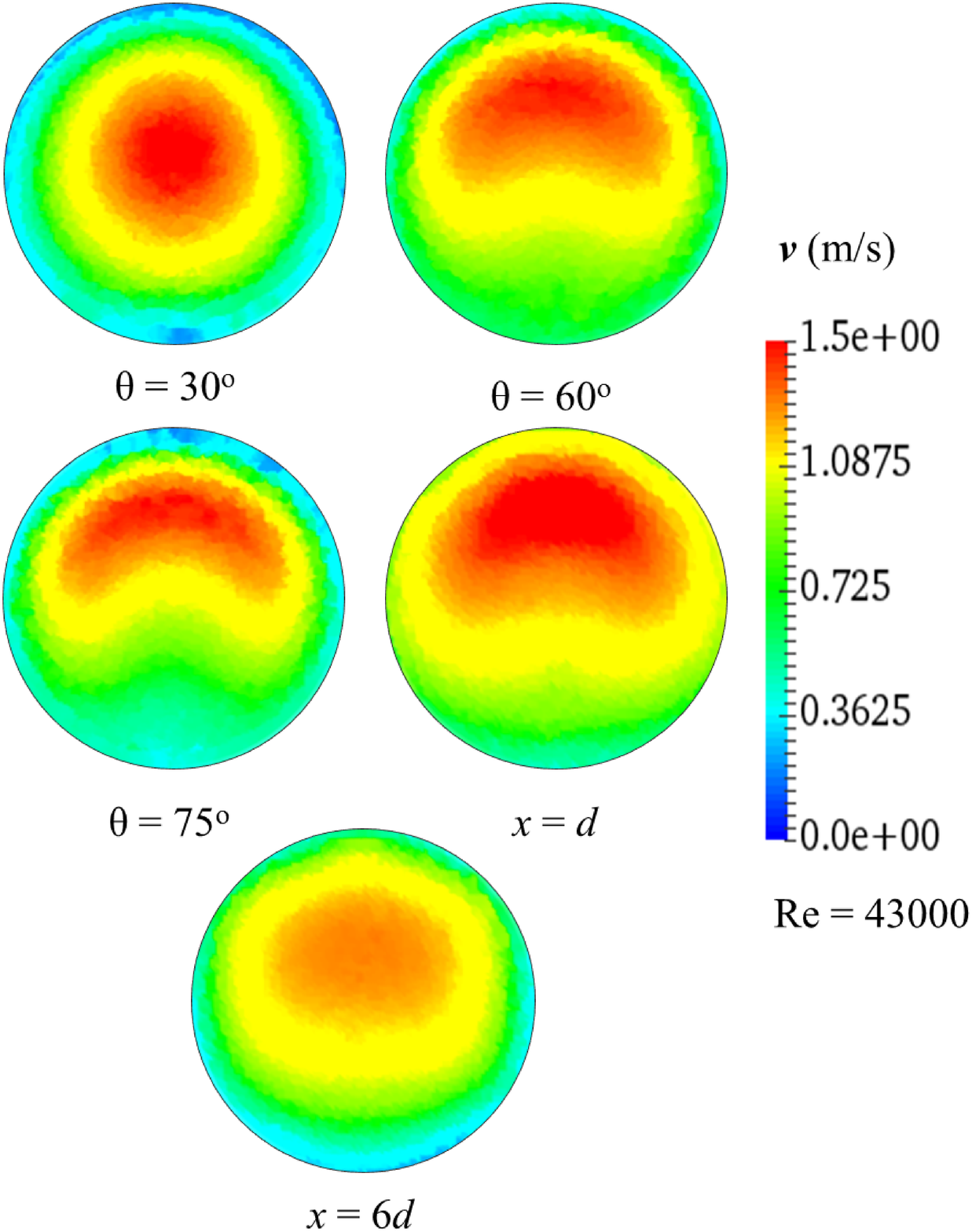,width=5.45in}}
\label{fig5}
FIGURE 5
\end{figure*}

\clearpage

\begin{figure*}
\centerline{\epsfig{figure=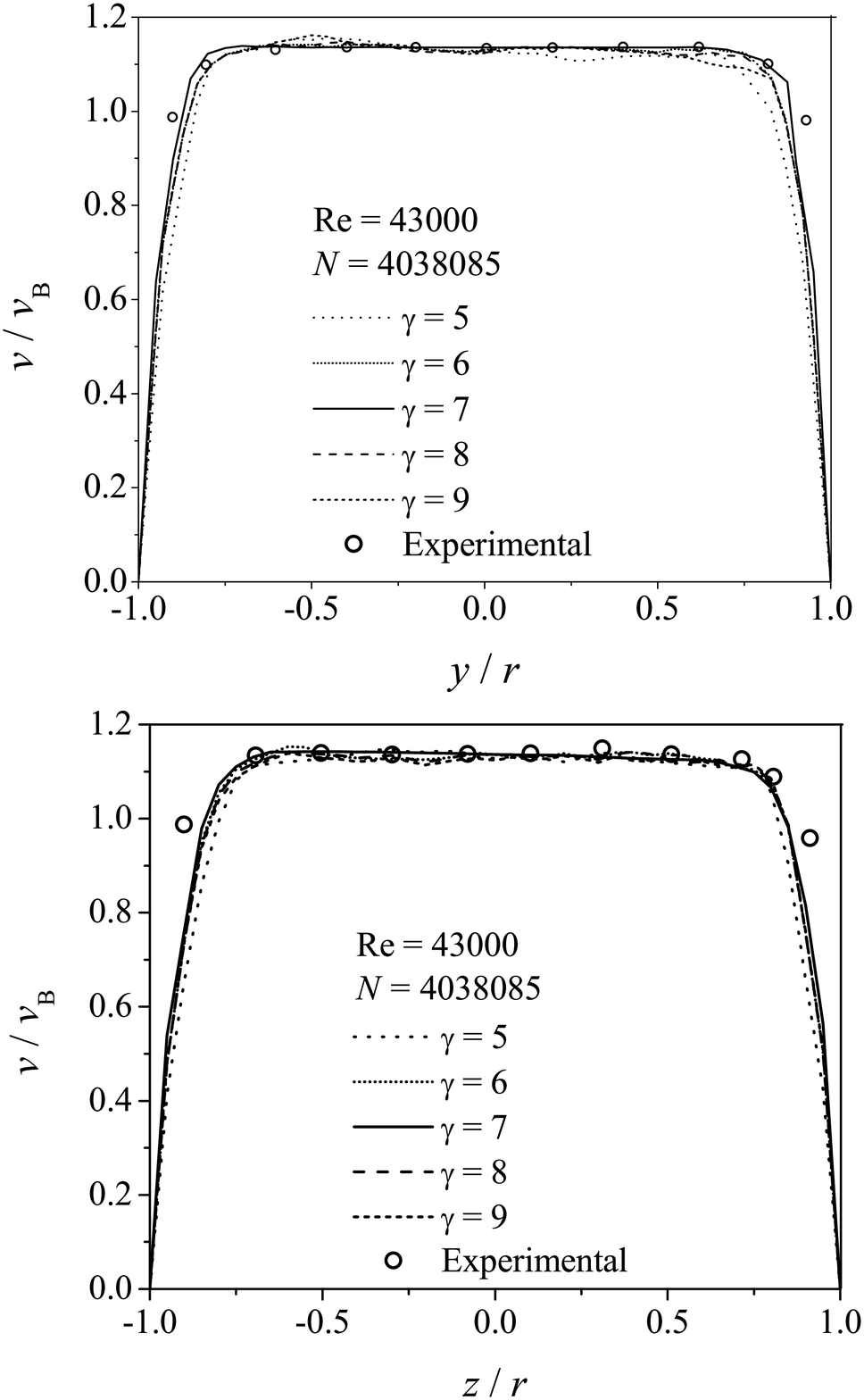,width=4.0in}}
\label{fig6}
FIGURE 6
\end{figure*}

\clearpage

\begin{figure*}
\centerline{\epsfig{figure=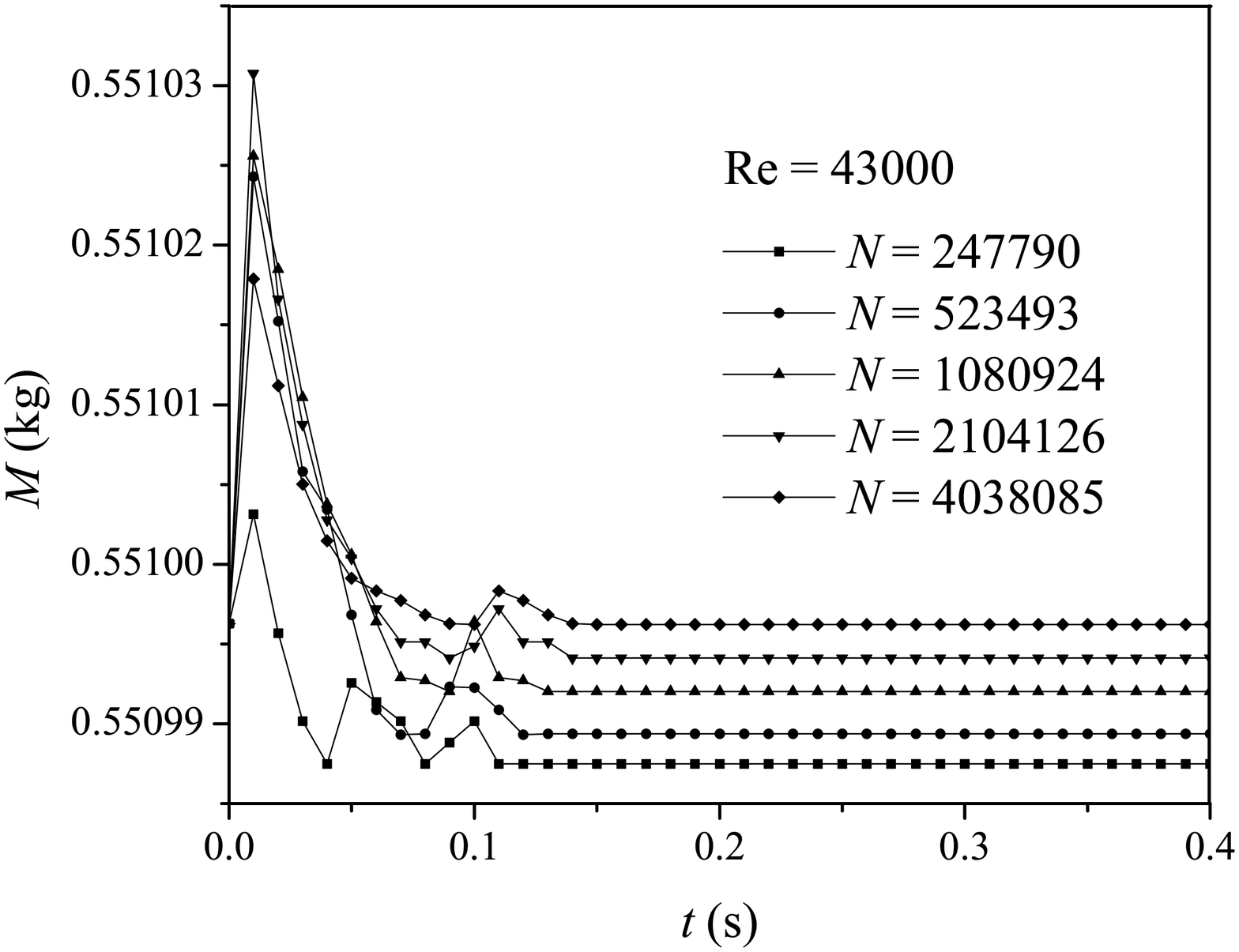,width=6.0in}}
\label{fig7}
FIGURE 7
\end{figure*}

\clearpage

\begin{figure*}
\centerline{\epsfig{figure=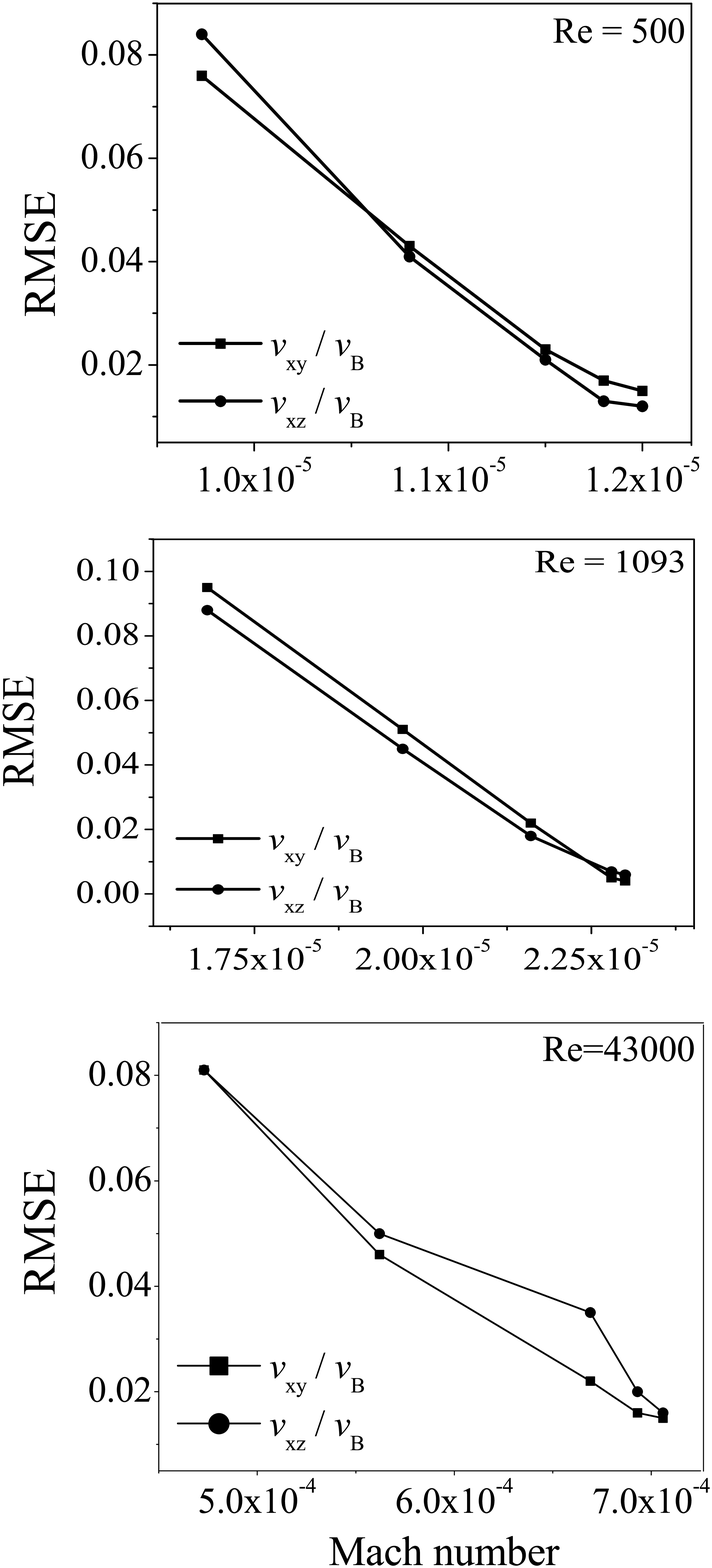,width=3.0in}}
\label{fig8}
FIGURE 8
\end{figure*}

\clearpage

\begin{figure*}
\centerline{\epsfig{figure=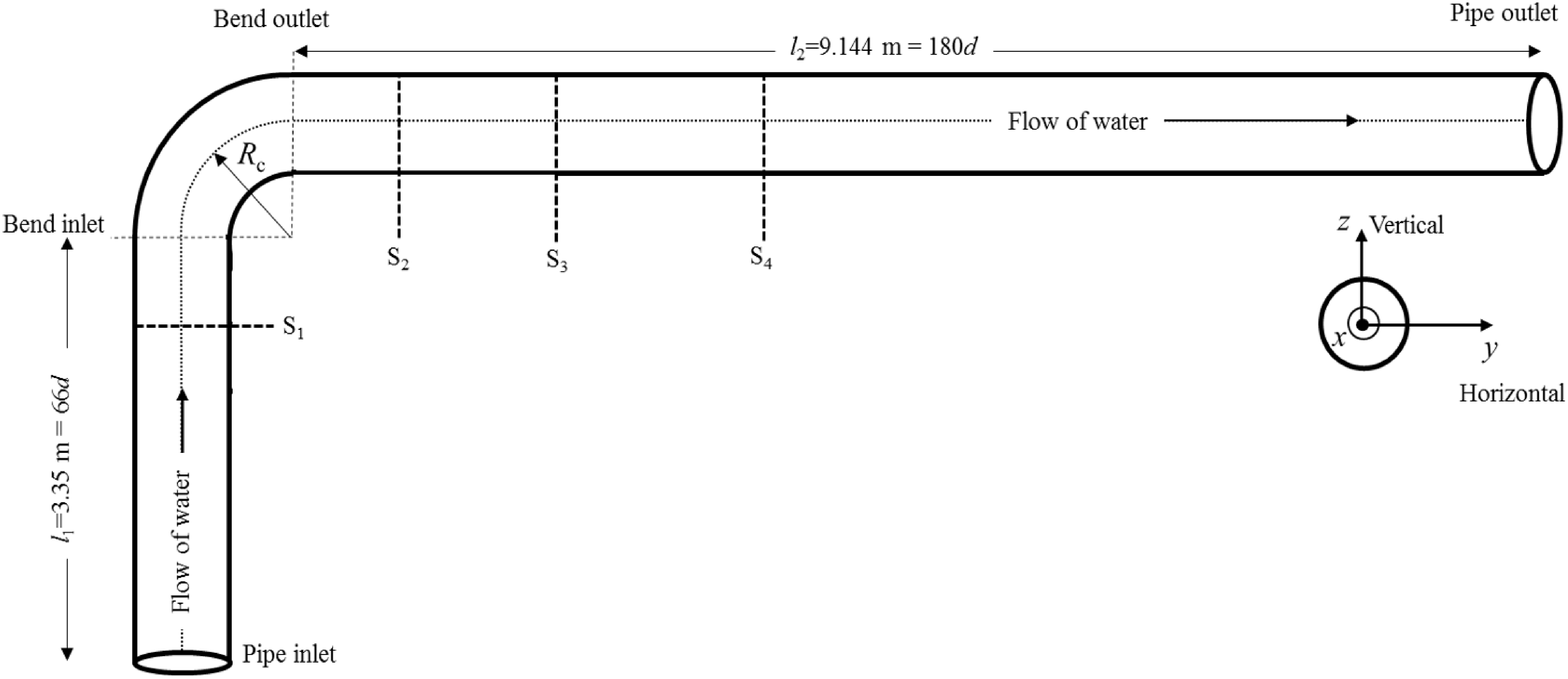,width=8.0in}}
\label{fig9}
FIGURE 9
\end{figure*}

\clearpage

\begin{figure*}
\centerline{\epsfig{figure=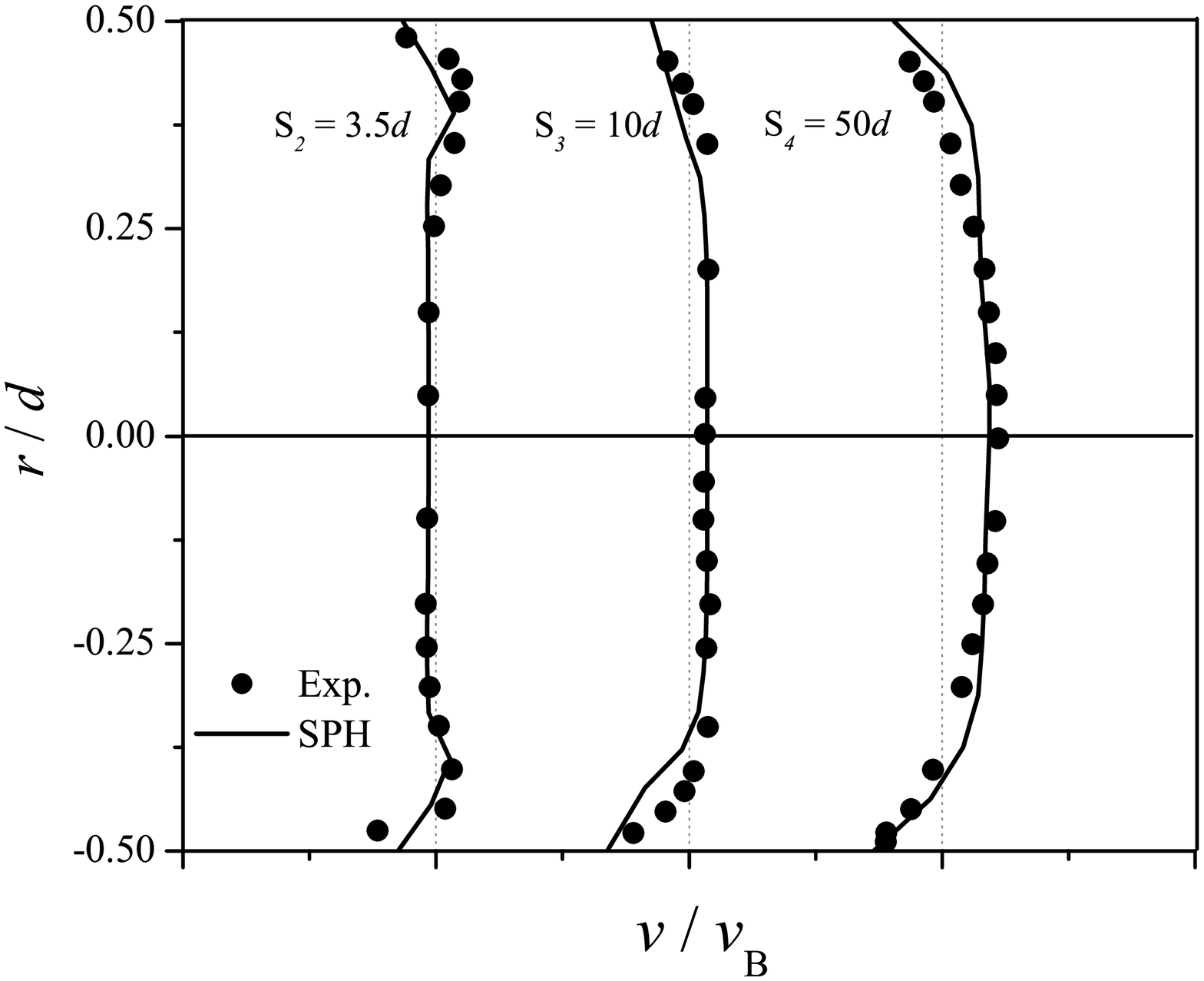,width=6.0in}}
\label{fig10}
FIGURE 10
\end{figure*}

\clearpage

\begin{figure*}
\centerline{\epsfig{figure=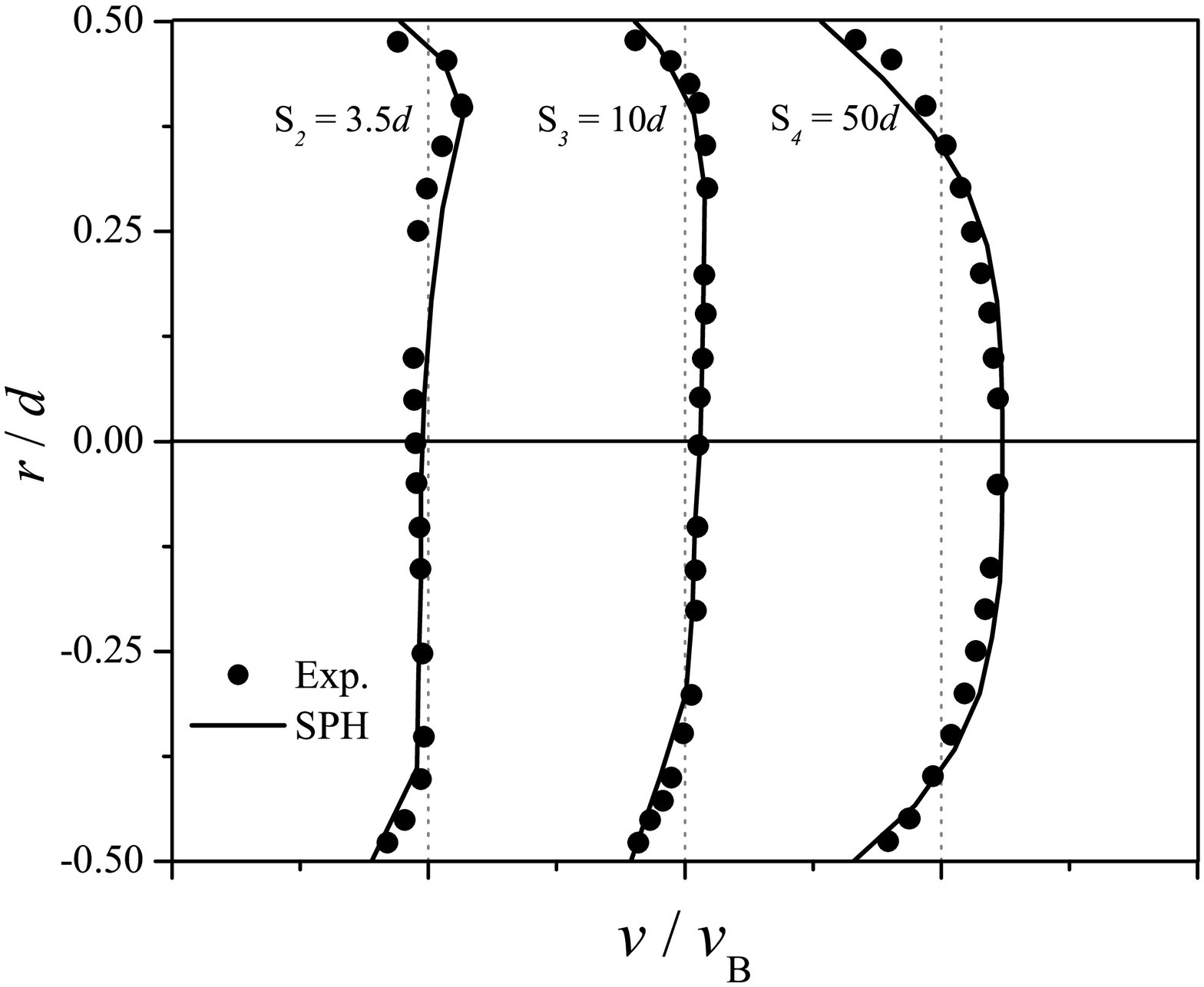,width=6.0in}}
\label{fig11}
FIGURE 11
\end{figure*}

\clearpage

\begin{figure*}
\centerline{\epsfig{figure=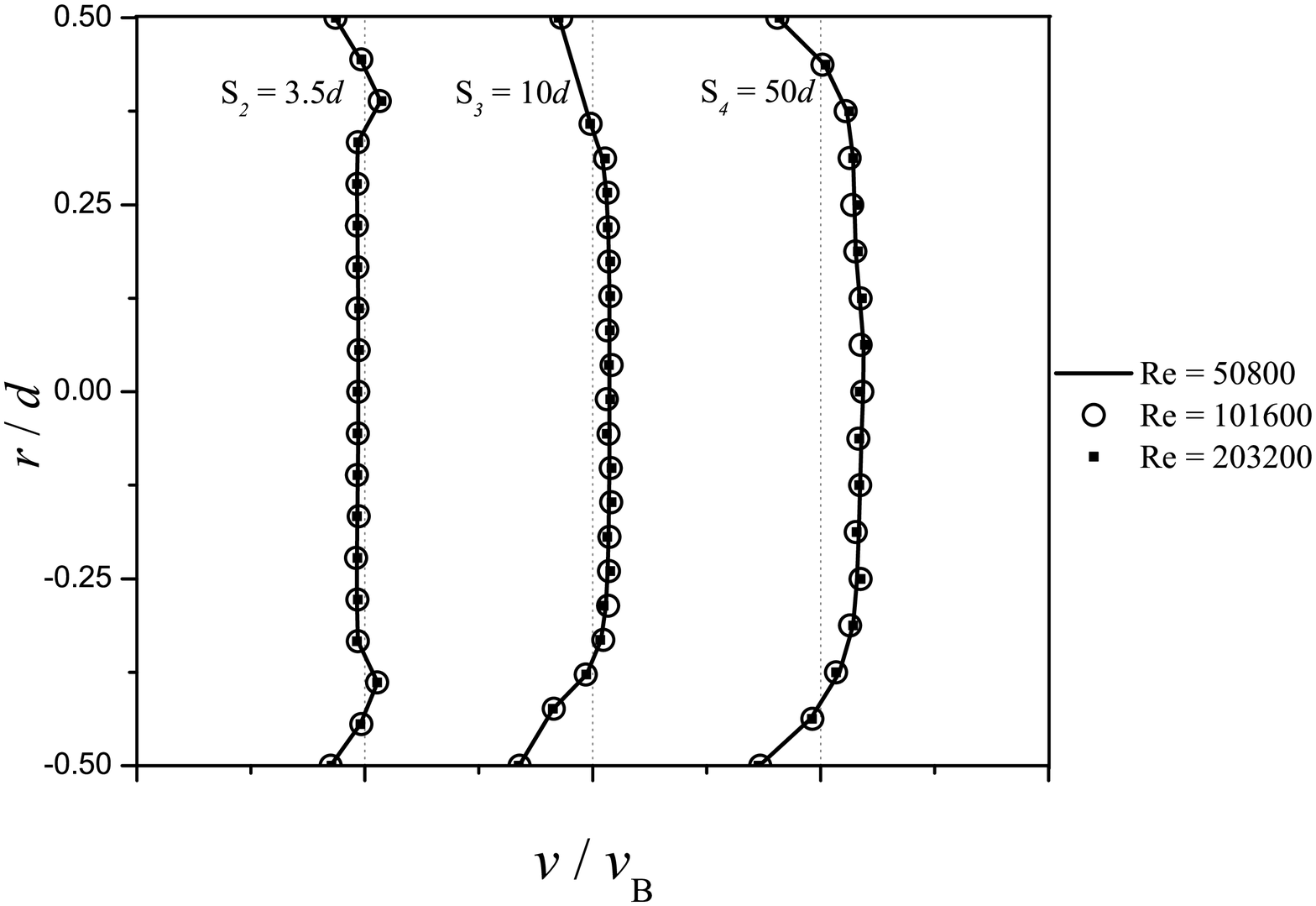,width=6.0in}}
\label{fig12}
FIGURE 12
\end{figure*}

\clearpage

\begin{figure*}
\centerline{\epsfig{figure=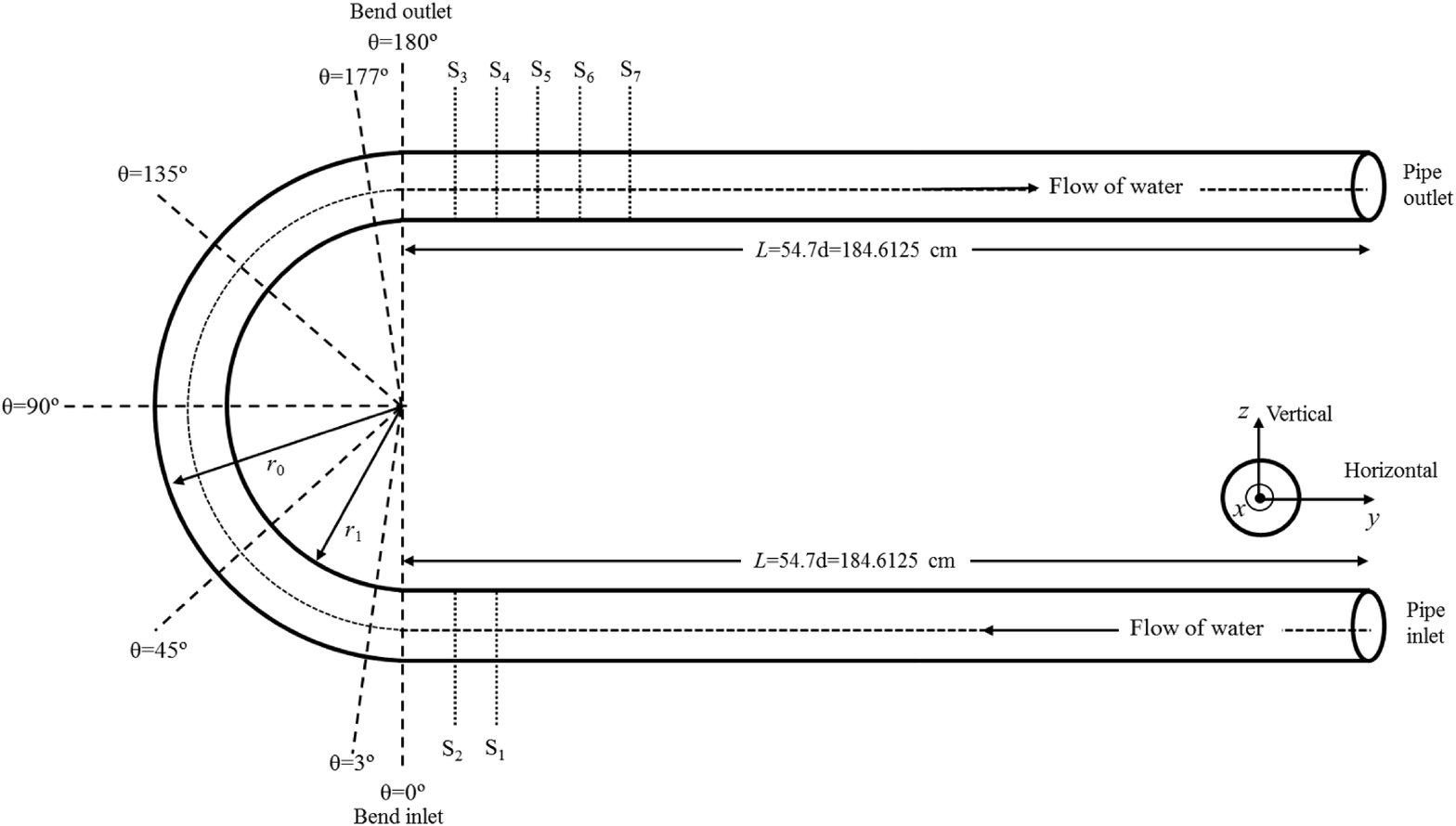,width=8.0in}}
\label{fig13}
FIGURE 13
\end{figure*}

\clearpage

\begin{figure*}
\centerline{\epsfig{figure=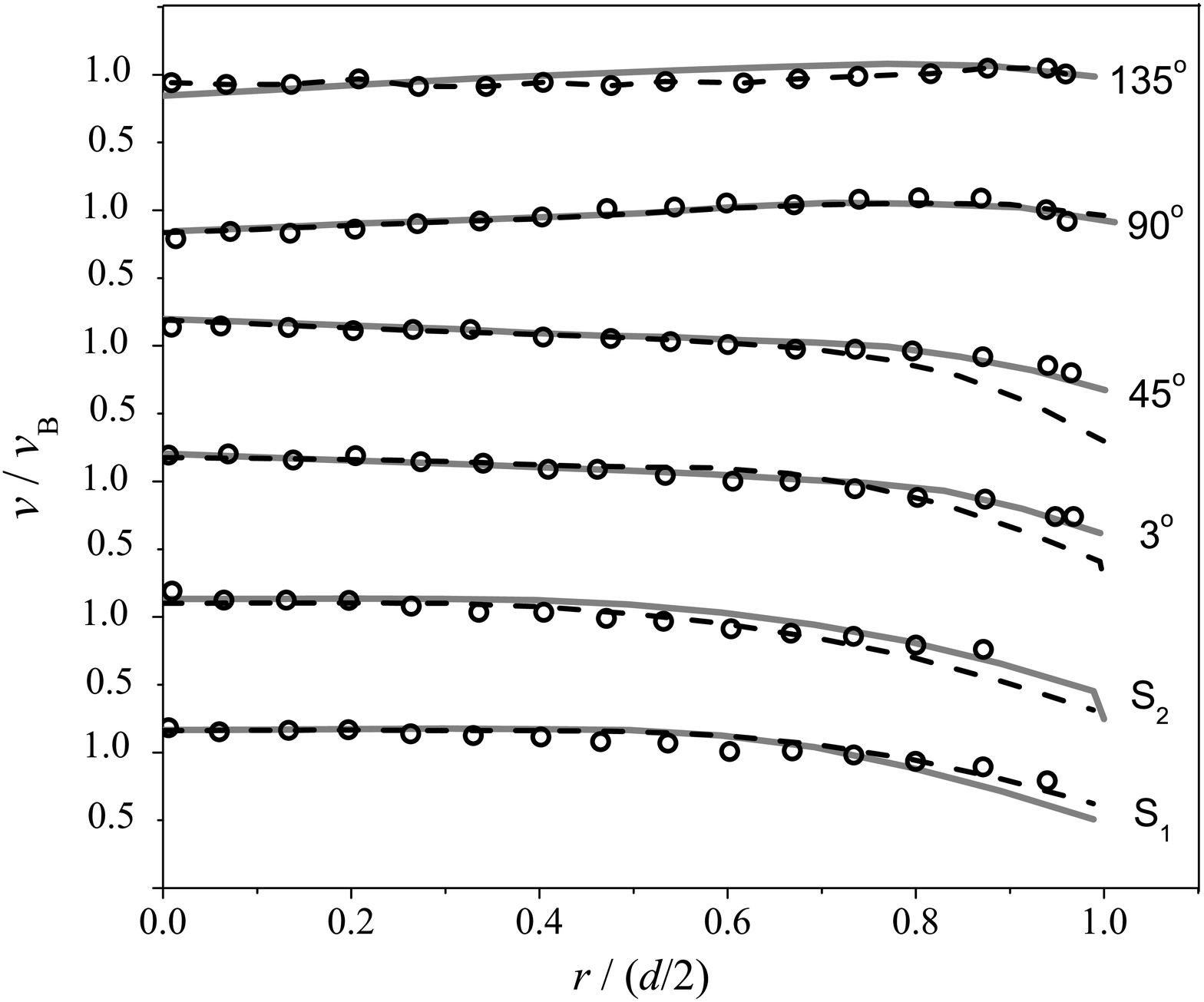,width=5.45in}}
\centerline{\epsfig{figure=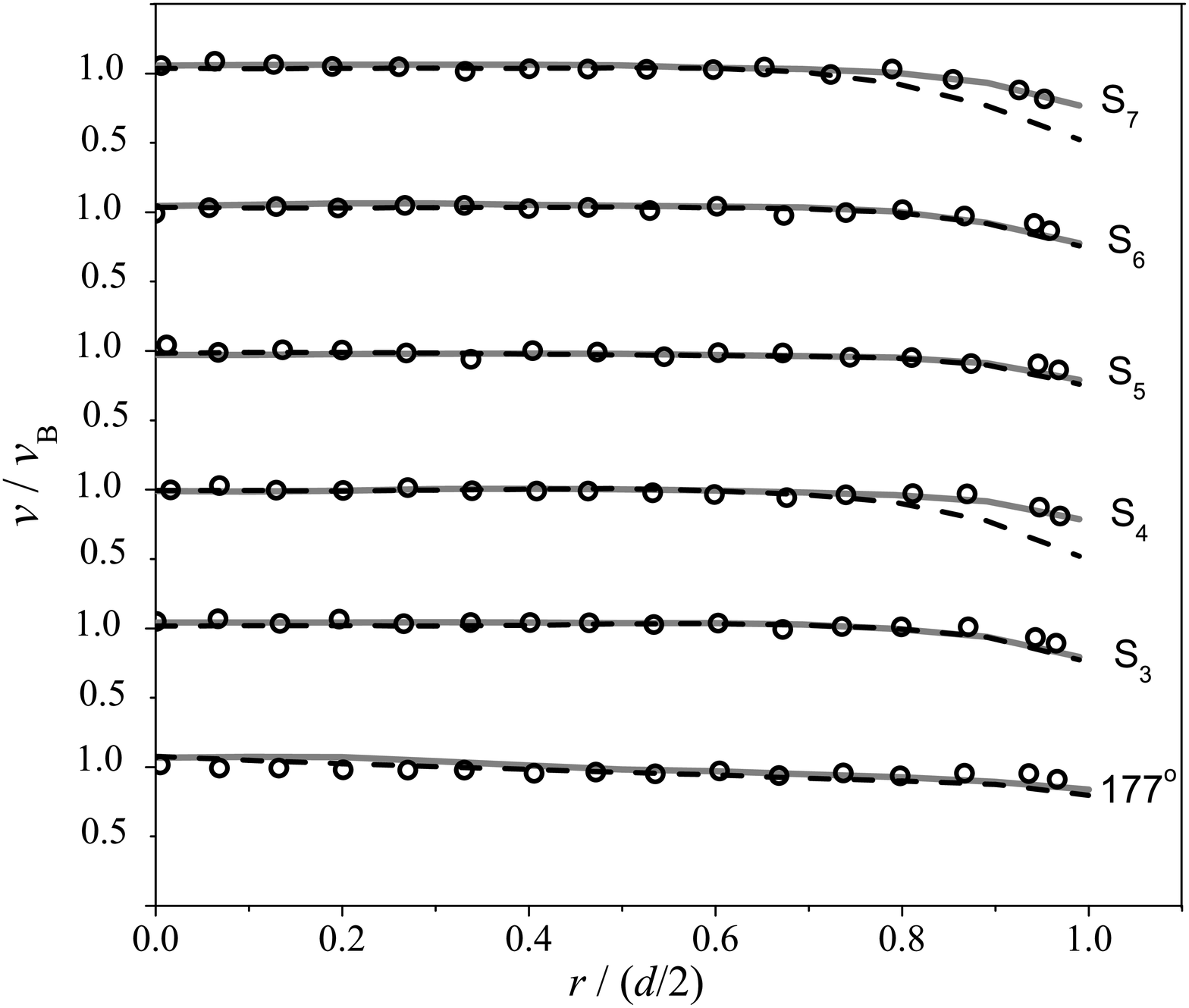,width=5.45in}}
\label{fig14}
FIGURE 14
\end{figure*}

\clearpage

\begin{figure*}
\centerline{\epsfig{figure=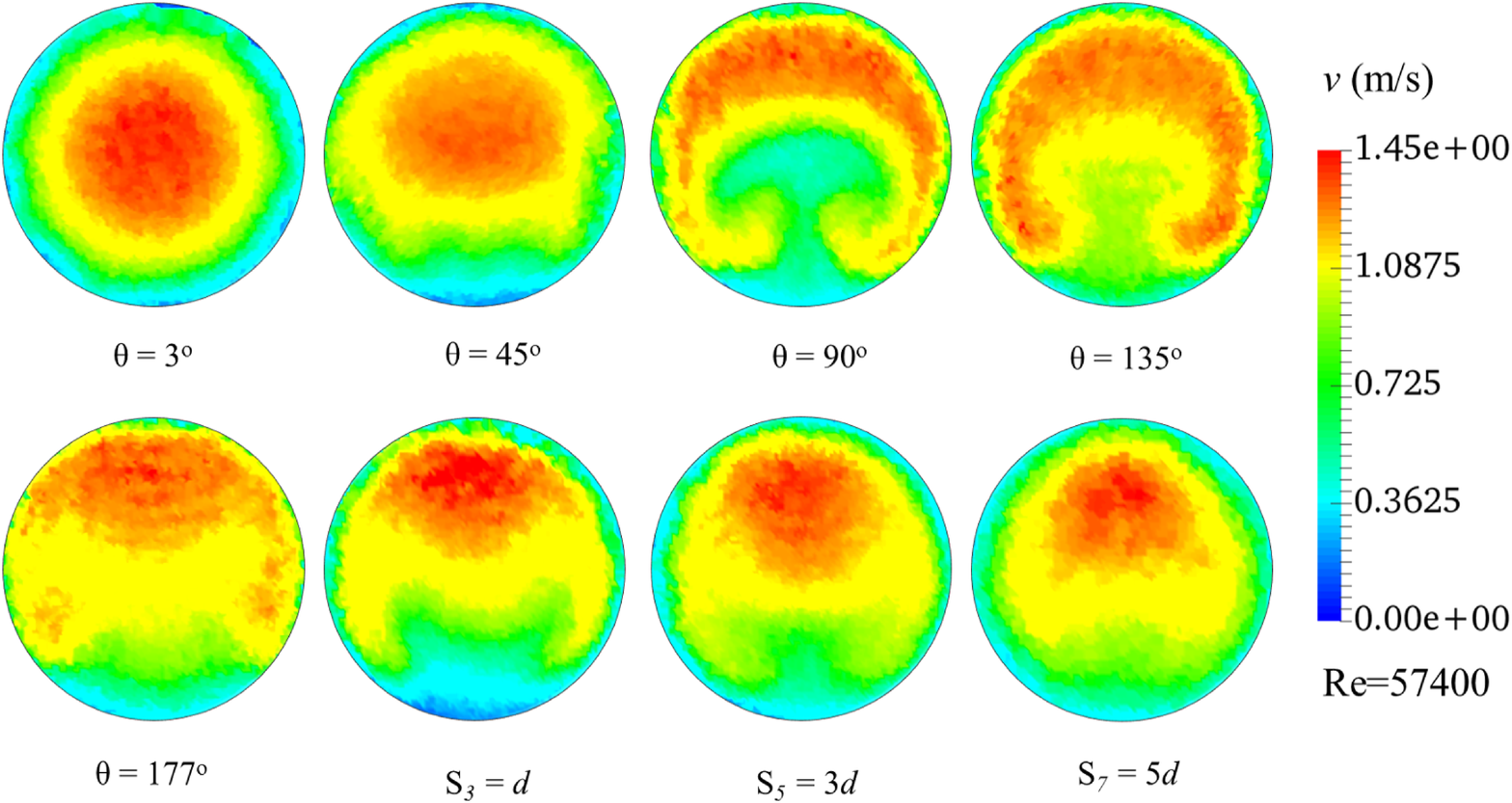,width=6.0in}}
\label{fig15}
FIGURE 15
\end{figure*}

\clearpage

\begin{figure*}
\centerline{\epsfig{figure=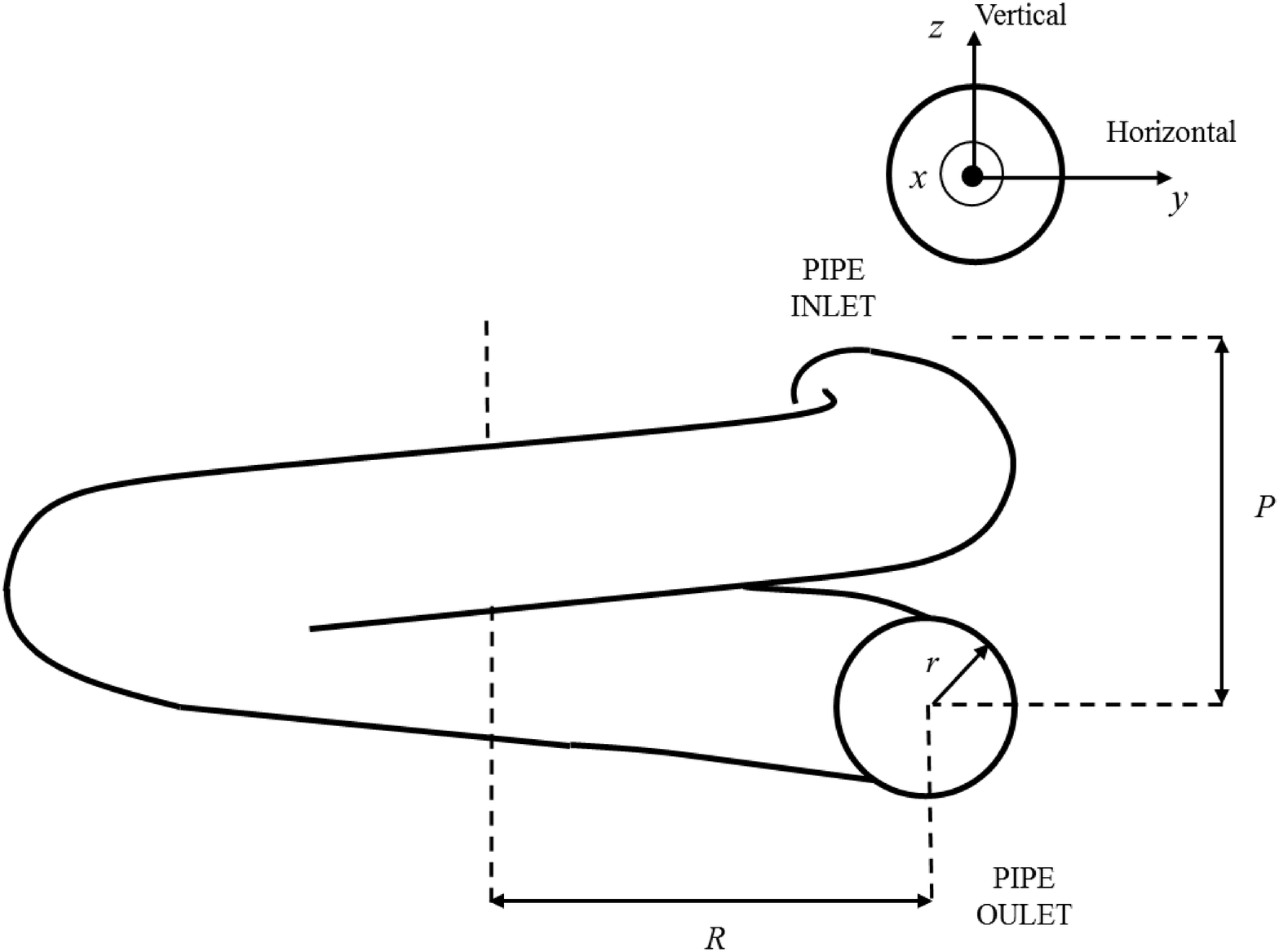,width=8.0in}}
\label{fig16}
FIGURE 16
\end{figure*}

\clearpage

\begin{figure*}
\centerline{\epsfig{figure=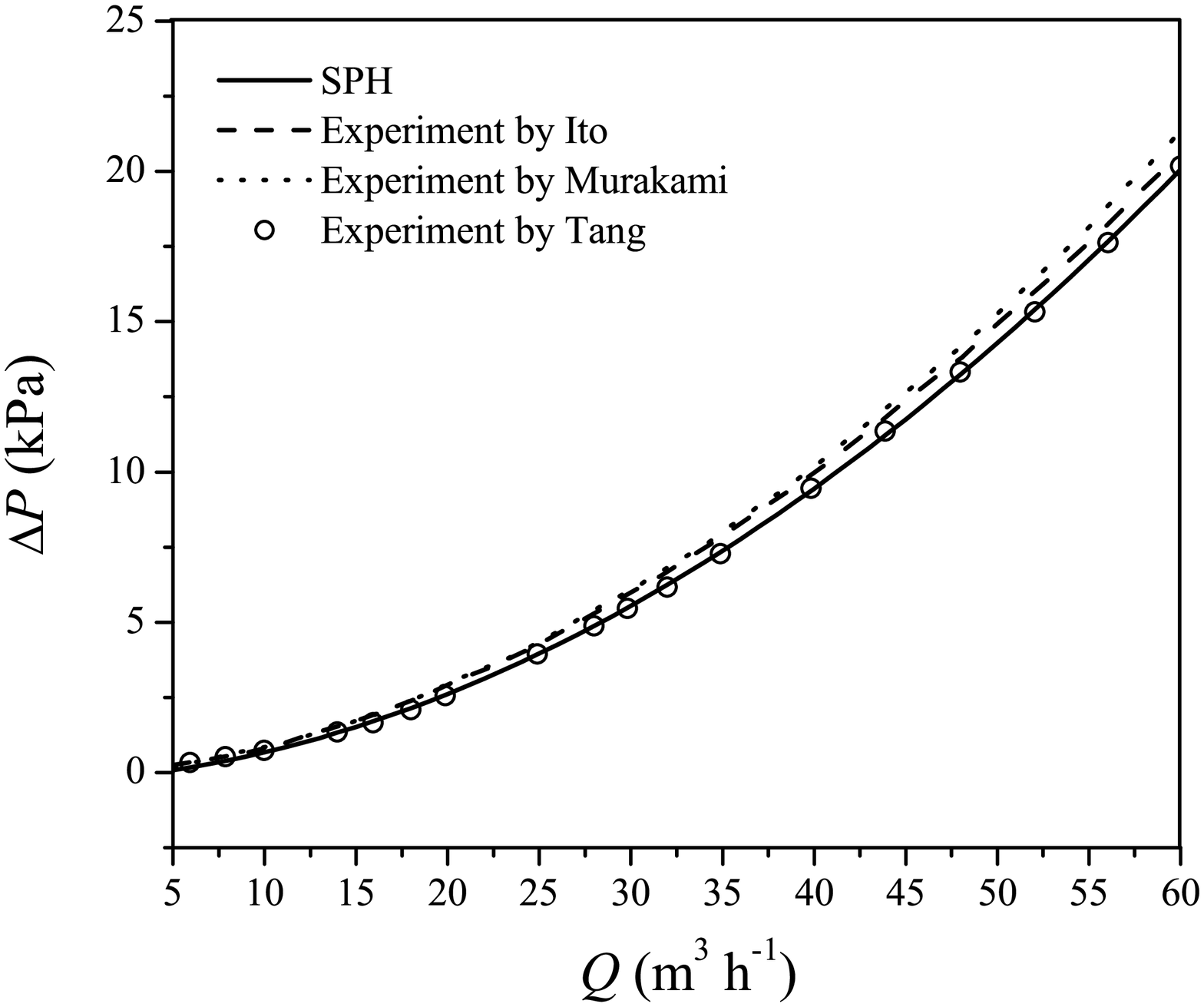,width=7.0in}}
\label{fig17}
FIGURE 17
\end{figure*}

\clearpage

\begin{figure*}
\centerline{\epsfig{figure=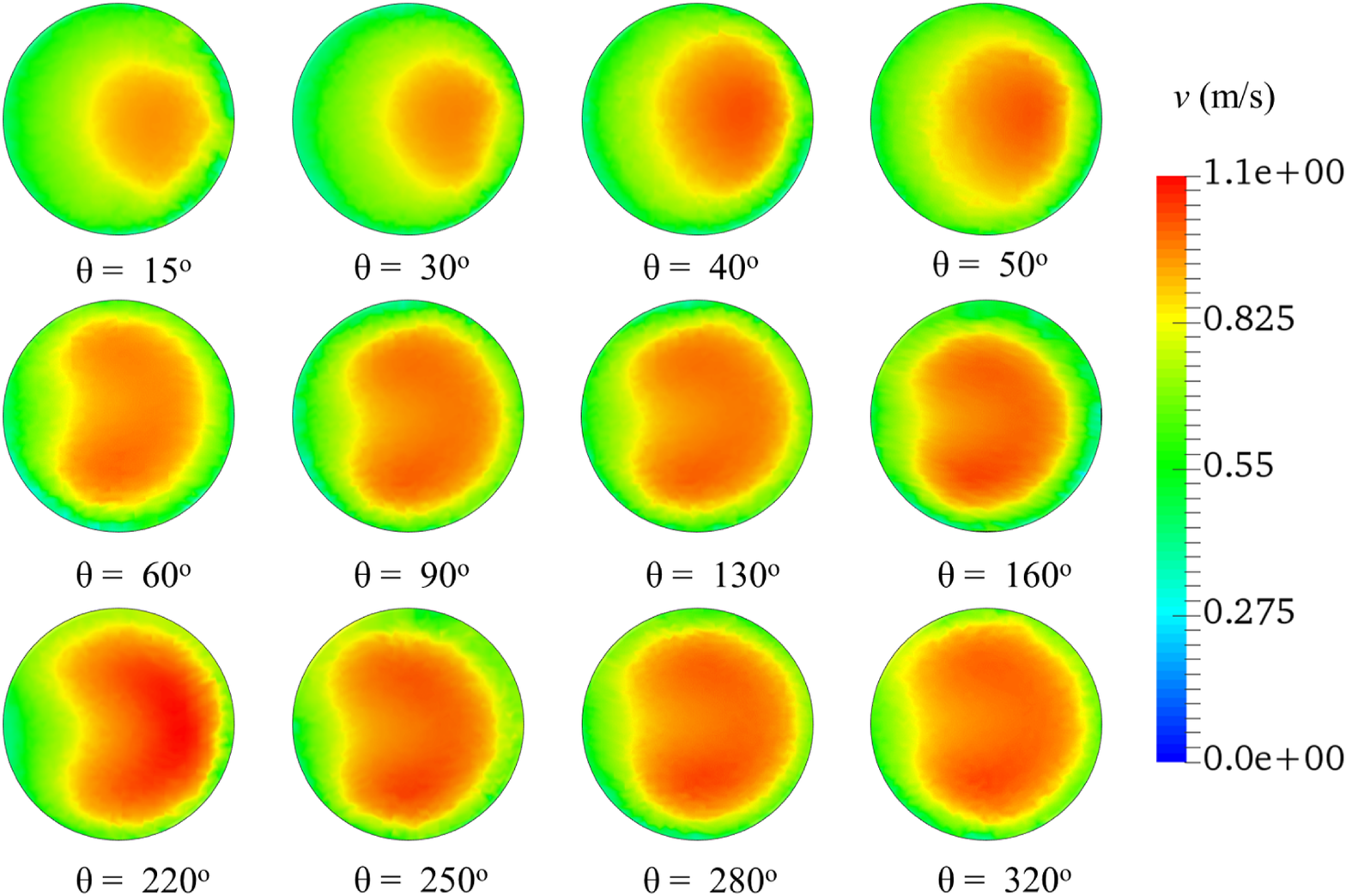,width=6.0in}}
\label{fig18}
FIGURE 18
\end{figure*}

\clearpage

\begin{figure*}
\centerline{\epsfig{figure=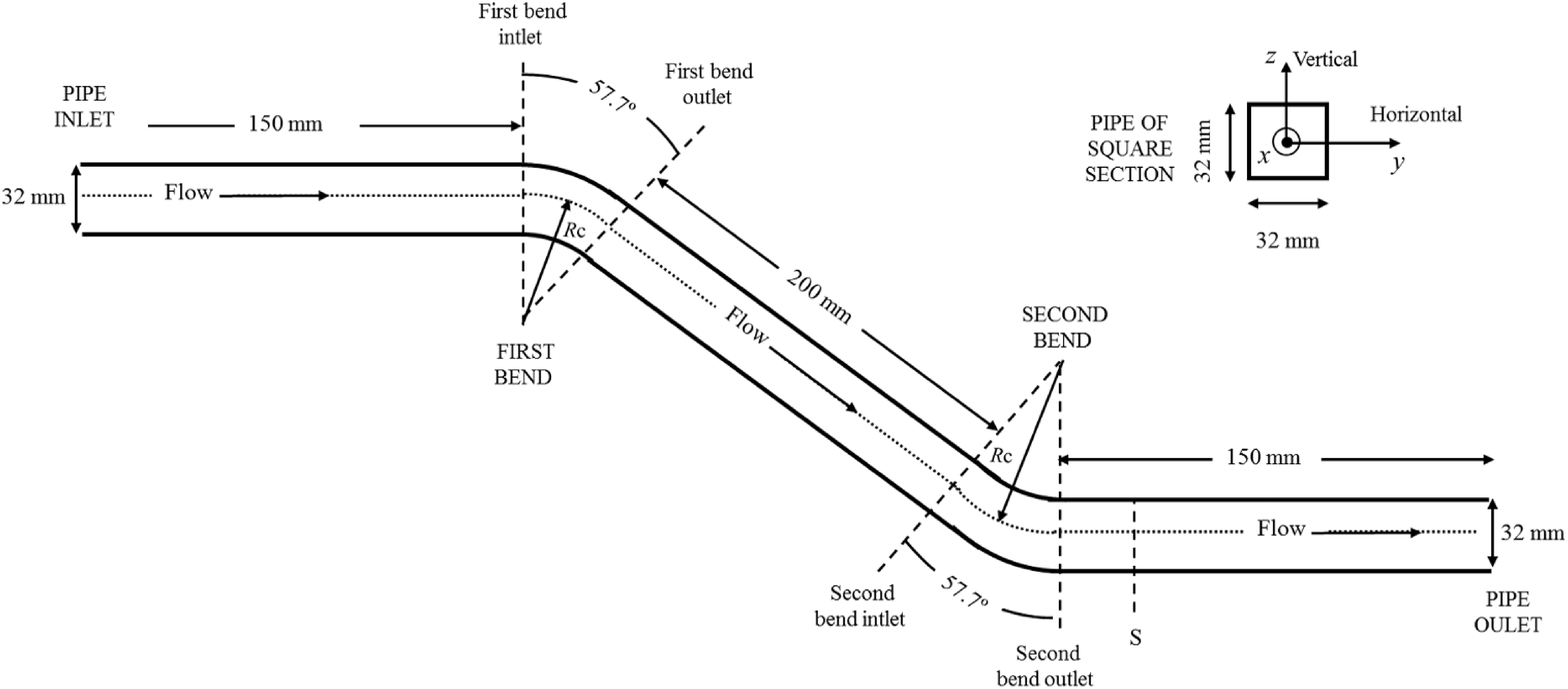,width=8.0in}}
\label{fig19}
FIGURE 19
\end{figure*}

\clearpage

\begin{figure*}
\centerline{\epsfig{figure=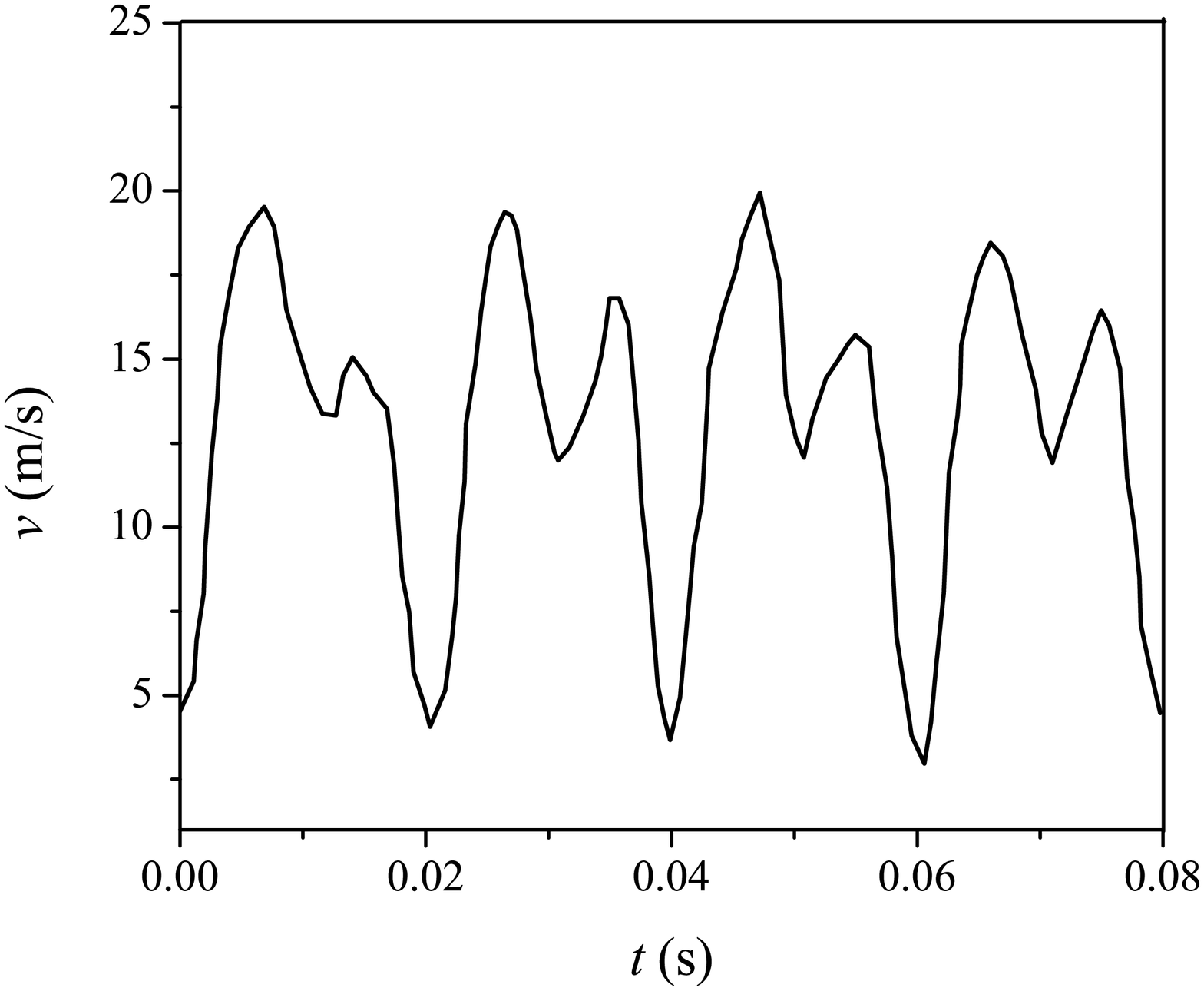,width=7.0in}}
\label{fig20}
FIGURE 20
\end{figure*}

\clearpage

\begin{figure*}
\centerline{\epsfig{figure=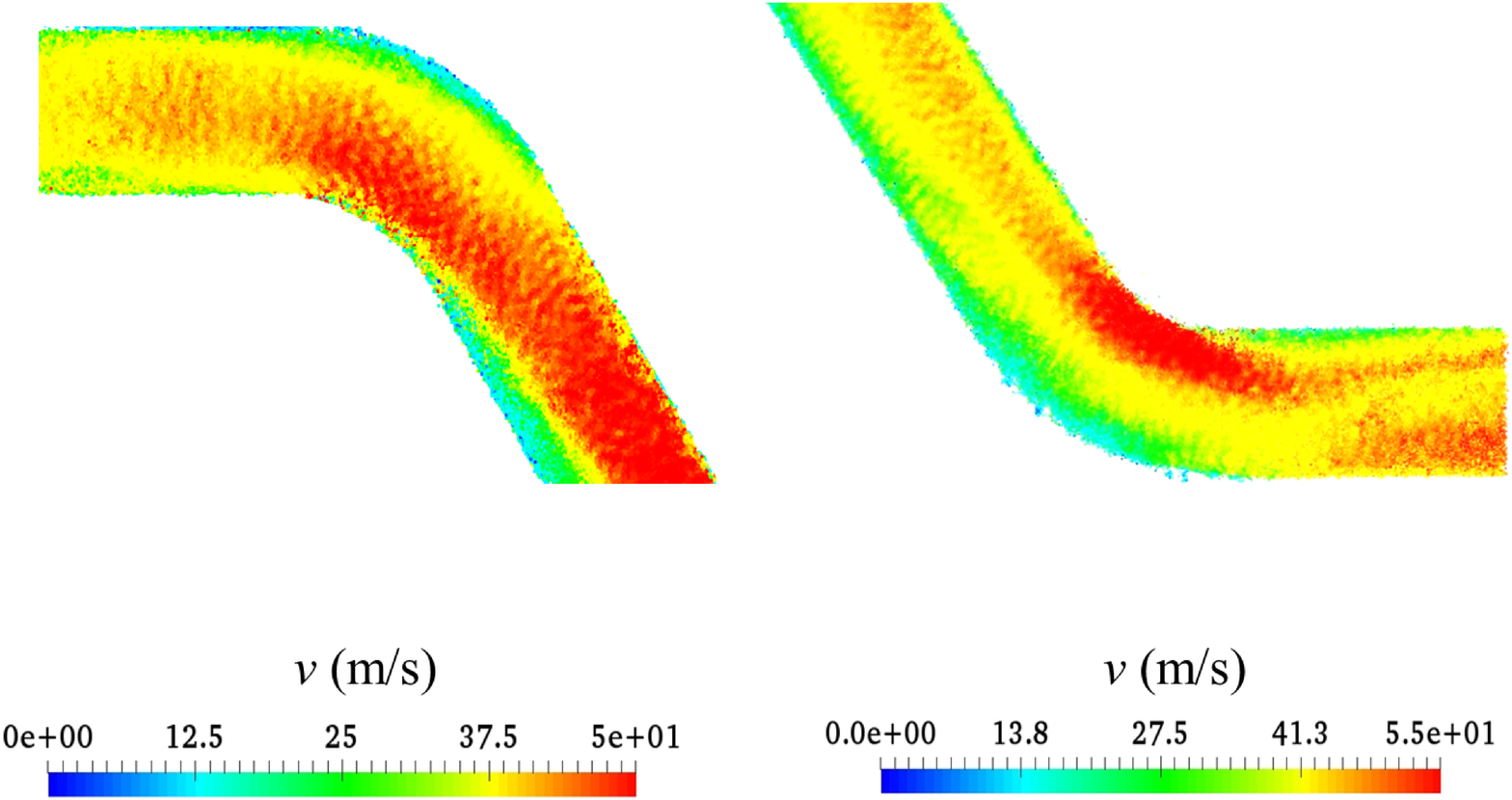,width=8.0in}}
\label{fig21}
FIGURE 21
\end{figure*}

\clearpage

\begin{figure*}
\centerline{\epsfig{figure=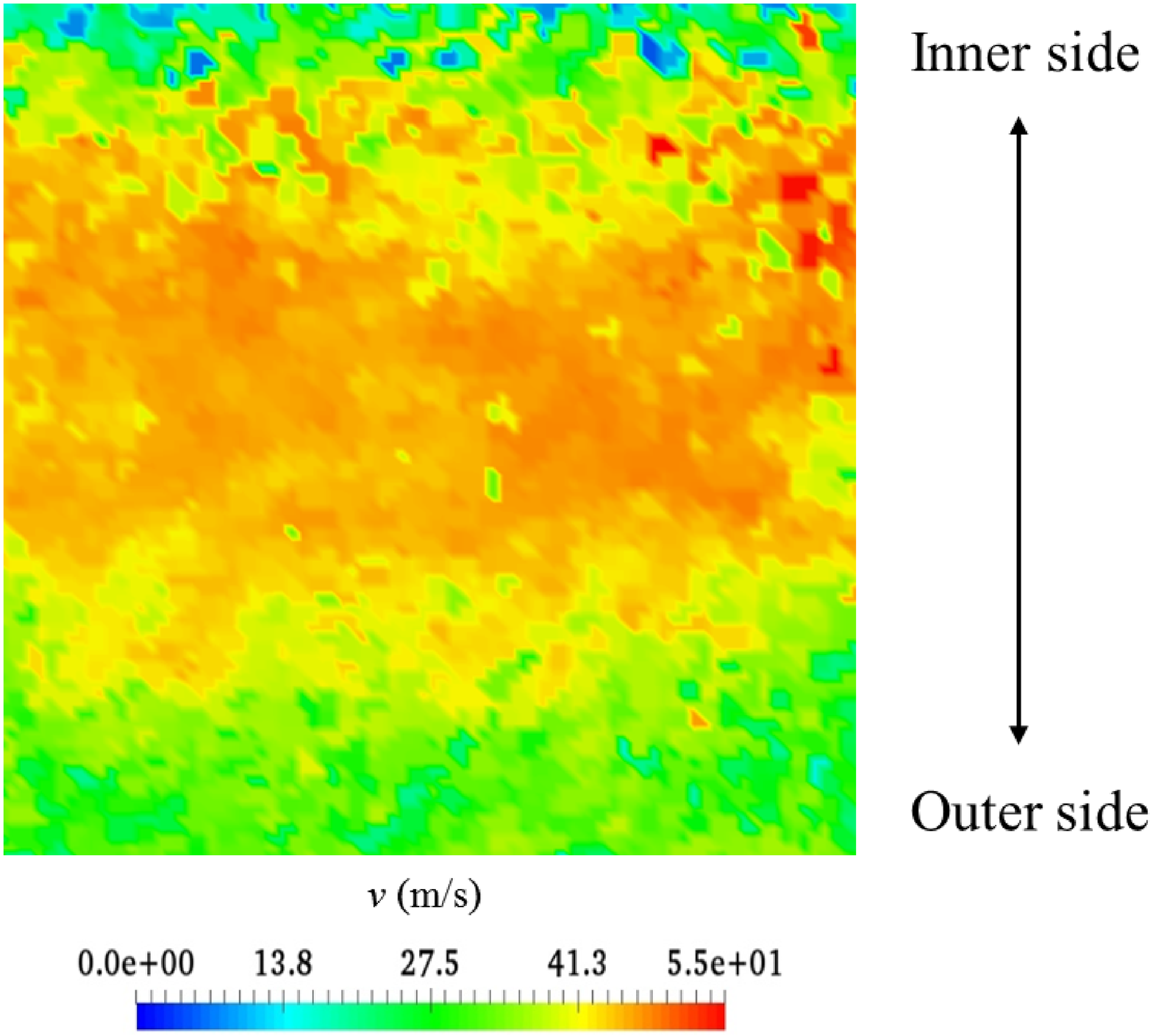,width=8.0in}}
\label{fig22}
FIGURE 22
\end{figure*}

\clearpage

\begin{figure*}
\centerline{\epsfig{figure=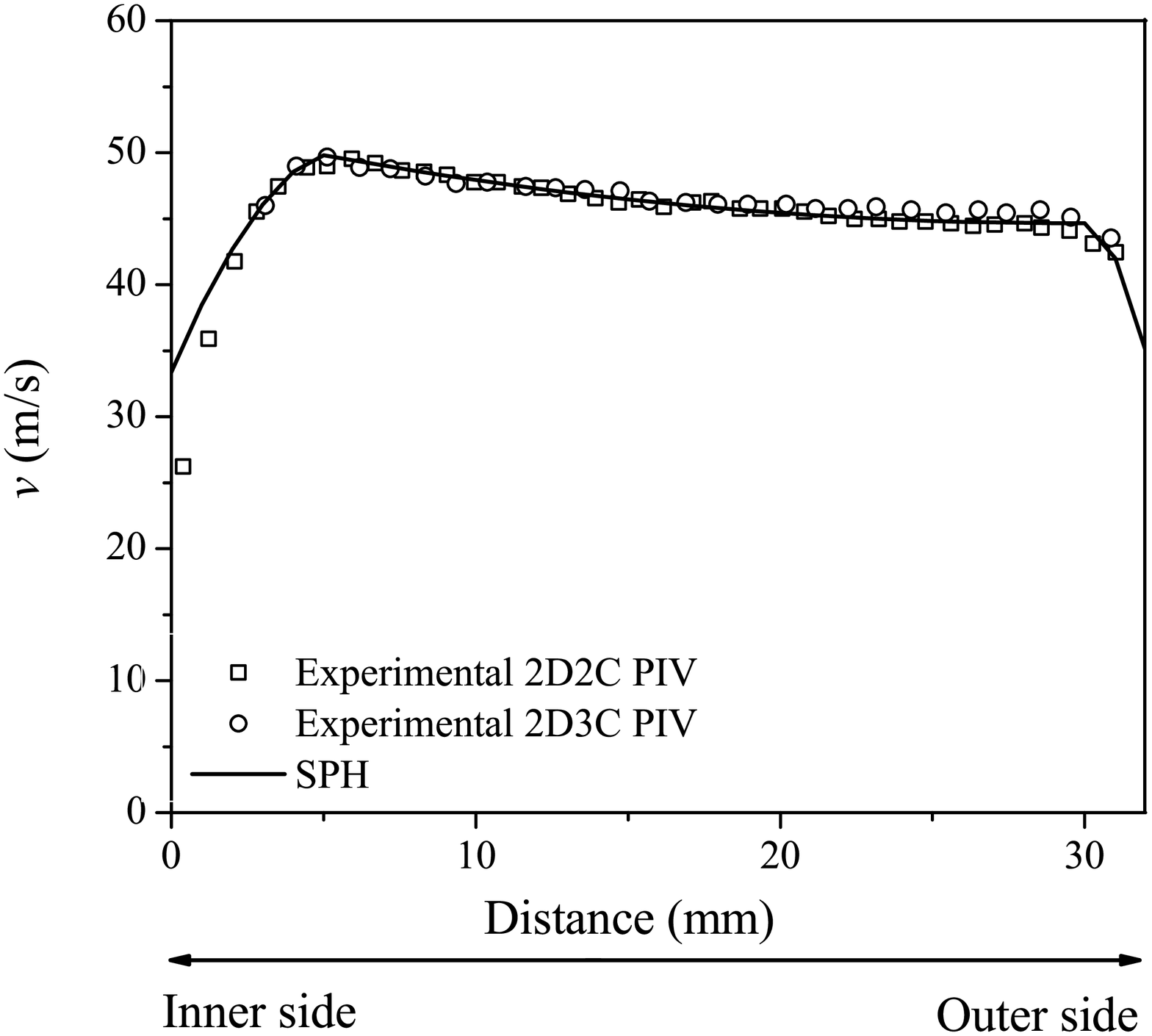,width=7.0in}}
\label{fig23}
FIGURE 23
\end{figure*}

\clearpage

\begin{figure*}
\centerline{\epsfig{figure=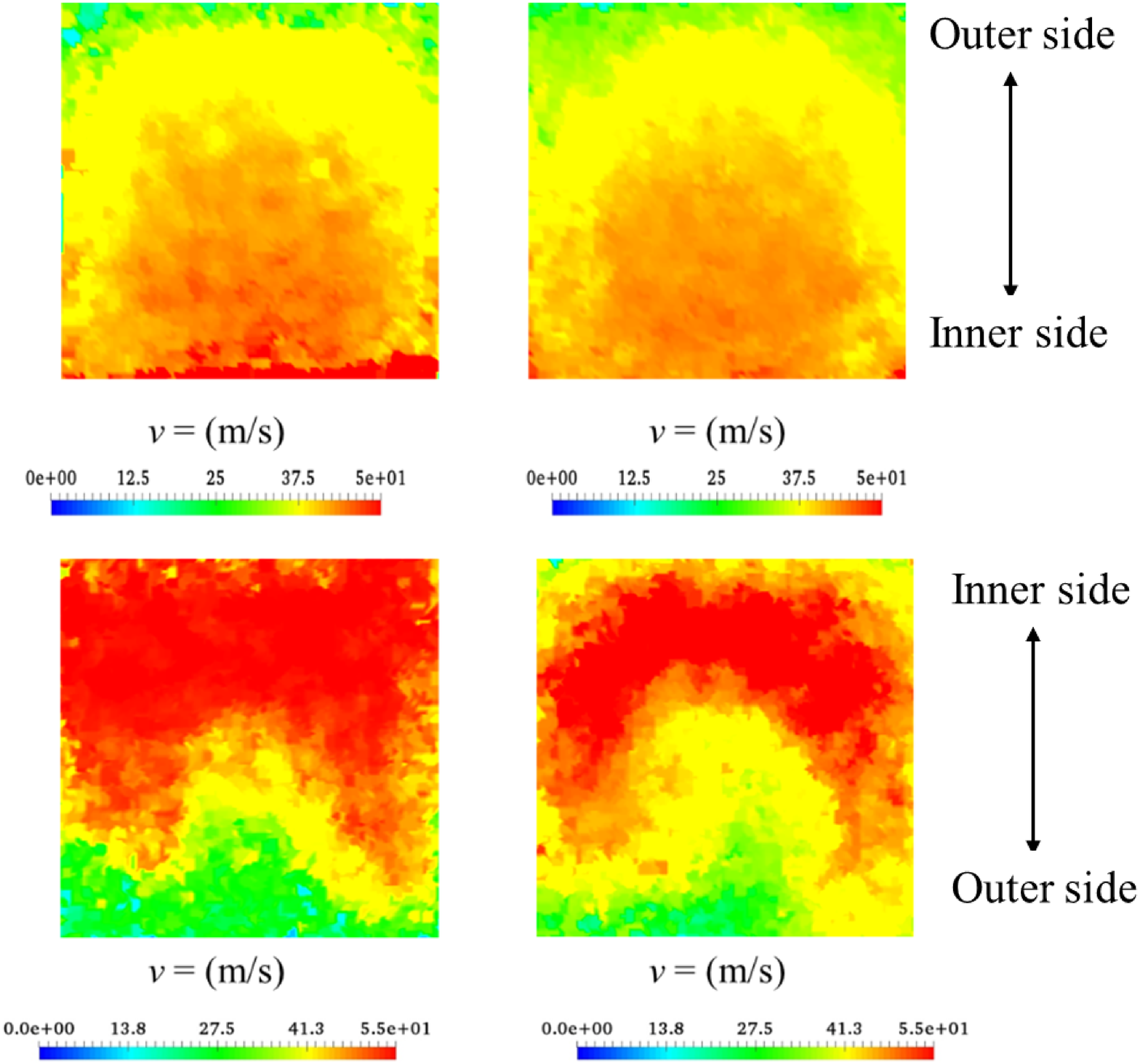,width=9.0in}}
\label{fig24}
FIGURE 24
\end{figure*}

\end{document}